\documentclass[twocolumn]{revtex4-2} 
\usepackage{amsmath,xfrac,amssymb,upgreek}
\usepackage[]{color}
\usepackage{multirow}
\usepackage{array}
\usepackage{microtype}
\usepackage[integrals]{wasysym}
\usepackage{placeins}
\usepackage{graphicx}

\newcommand{\figref}[1]{{\color{blue}Figure\,\ref{#1}}} 
\newcommand{\tabref}[1]{{\color{blue}Table\,\ref{#1}}} 
\newcommand{\secref}[1]{{\color{blue}Section\,\ref{#1}}} 
\newcommand{\bistI}{\!\!\; }

\newcommand{\distIII}{\, }
\newcommand{\InEqDist}{\distIII}

\newcommand{\lt}{\InEqDist{<}\InEqDist}
\newcommand{\gt}{\InEqDist{>}\InEqDist}
\newcommand{\br}[1]{\left(#1\right)}
\newcommand{\sq}[1]{\left[{#1}\right]}
\newcommand{\ud}[1]{\mathrm{d}{#1}}
\renewcommand{\eqref}[1]{{\color{blue}Eq.\ref{#1}}}
\newcommand{\eexp}[1]{\mathrm{e}^{#1}}
\newcommand{\Tud}[3]{#1^{#2}_{\phantom{#2}#3}}
\newcommand{\ydens}{\uppsi}
\renewcommand{\theta}{\upvartheta}
\renewcommand{\alpha}{\upalpha}
\renewcommand{\beta}{\upbeta}
\renewcommand{\upsilon}{\upupsilon}
\renewcommand{\chi}{\upchi}
\renewcommand{\kappa}{\upkappa}
\renewcommand{\epsilon}{\upepsilon}
\renewcommand{\tau}{\uptau}
\newcommand{\IDc}{\bistI c}
\newcommand{\uc}{u_{\IDc}}
\newcommand{\xc}{x_{\IDc}}
\newcommand{\yc}{\ydens_{\IDc}}
\newcommand{\bc}{\beta_{\IDc}}
\newcommand{\kc}{\kappa_{\IDc}}
\newcommand{\au}{\alpha_{\bistI\upsilon}}

\usepackage{xspace} 
\definecolor{mygreen}{RGB}{30,150,30} 
\definecolor{mygrey}{RGB}{125,125,125}

\newcommand{\alpfty}{\gamma_{\!\infty}}
\newcommand{\rfty}{r_{\!\infty}}
\newcommand{\bfty}{\beta_{\!\infty}}
\newcommand{\dfty}{\gamma_{\!\infty}}
\newcommand{\nfty}{n_{\!\infty}}
\newcommand{\tfty}{\theta_{\!\infty}}
\newcommand{\kapfty}{\kappa_{\!\infty}}
\newcommand{\epsfty}{\epsilon_{\!\infty}}
\newcommand{\yfty}{\ydens_{\!\infty}}

\DeclareMathOperator{\sgn}{sgn}
\newcommand{\Kappa}{K}    
\newcommand{\Epsilon}{E}     
       
\newcommand{\alen}{a}
\newcommand{\blen}{b}
\newcommand{\Qupsilon}{Q}

\usepackage{hyperref}

\begin{document}

\title{Cylindrically symmetric  
radial accretion onto Levi-Civita string-like source}

\author{{\L}ukasz Bratek}
\email{Lukasz.Bratek@pk.edu.pl}
\affiliation{Institute of Physics, Cracow University of Technology, ul.~Podchor\c{a}\.{z}ych 1, PL-30084 Krak\'{o}w, Poland}

\author{Joanna Ja{\l}ocha}
\affiliation{Institute of Physics, Cracow University of Technology, ul.~Podchor\c{a}\.{z}ych 1, PL-30084 Krak\'{o}w, Poland}

\author{Marek Kutschera}
\affiliation{Institute of Physics, Jagiellonian University, PL-30059  Krak\'{o}w, Poland}

\date{\today}

\begin{abstract} 
Radial steady-state accretion of polytropic matter is investigated under cylindrical symmetry in the Levi-Civita background metric. The model can be considered as a cylindrical analog of Bondi accretion in strong gravitational field. As a byproduct of this study, the issue of defining the line mass density is addressed and the role of the metric free parameters is discussed on the example of physical observables. The form of radial accretion equations is insensitive to the structure of the interior solution. Accordingly, the accretion solution analysis can be limited to a special Wilson form of Levi-Civita metric describing a structureless homogeneous string. 
 \end{abstract} 


\maketitle

 
\section{Introduction}

The  accretion of gas onto elongated concentrations of matter -- such as filaments, arms or slender rings, which are modeled as infinite cylinders \citep{1964ApJ...140.1056O} -- can be approximately described in the field of such cylinders. 
Radial and spiraling non-relativistic  accretion of test matter was investigated in this context in the power-law background potentials \citep*{Bratek_2019}.\footnote{The motivation behind considering cylindrically
symmetric purely radial horizontal flows has been already discussed in \citep{Bratek_2019} were some references to the relevant astrophysical literature is also given.} As a limit within this class of models, one can distinguish radial accretion in the vacuum logarithmic potential. The model is noteworthy for it can be viewed as the cylindrical analogy of the \citet{1952MNRAS.112..195B} spherical accretion model. The Bondi model was later generalized by \citet{1972Ap&SS..15..153M} to describe radial accretion in close proximity of condensed objects. In similar lines, it is instructive to make a strong-field extension of the vacuum model under cylindrical symmetry.
Accordingly, the present paper investigates purely radial steady-state accretion of test fluid
in the cylindrically-symmetric and static spacetime of a line source. Similar to Bondi-Michel model, self-gravity and dissipative processes of the accreting matter is neglected.

The spacetime metric of interest can be viewed as the exterior solution outside a condensed thin cylinder or a string. Physically, the metric describes approximately the spacetime close to a massive elongated object in the same sense as the logarithmic potential describes an elongated axi-symmetric rod-like mass distribution in the leading approximation at short distances lower than the rod length. 

The paper is organized as follows. The next section begins with a short derivation of the required metric form. Next, the physical role of free  parameters of the metric is clarified on the example of redshift and Kepler problem. Then, the role of these parameters in the definition of the line mass density  is discussed in terms of Komar integral.  Next, the accretion equations in the Levi-Civita metric are derived and shown to be formally the same in the case of purely radial flow as for the special Wilson metric. Finally, the radial accretion solutions are studied and their physical discussion is given.

\section{\label{sec:spacetime}Remarks on Levi-Civita spacetime}
 The vacuum spacetime metric exterior to a static cylindrical body was found by \citet{Civita2019} a century ago. He described the solution to be interesting as it {\it `determined exactly the geometrical, mechanical and optical influence of a material cylinder'}. Later on, the solution was extensively studied and interpreted physically  \citep*[][references therein]{Bonnor_1991}, \citep*{Bonnor1999,Herrera2001}. 
\renewcommand{\Qupsilon}{\upsilon}  
A special form of Levi-Civita metric can be represented in the circumferential radius gauge as
\begin{equation}\label{eq:metric}\!\!\;
\ud{s^2}{=}{-\!}\br{\frac{r}{a}}^{\!2\Qupsilon}\!\!\!c^2{\ud{t}}^2{+\!}
  \br{\frac{r}{a}}^{\!\frac{2{\Qupsilon}^2}{1{+}\Qupsilon}}\!\!{\ud{r}}^2{\!+}  
  r^2{\ud{\phi}}^2{+\!}\br{\frac{a}{r}}^{\!\frac{2\Qupsilon}{1{+}\Qupsilon}} \!\!{\ud{z} }^2\!,\!\end{equation} with $\Qupsilon{>}0$ being a number and $a{>}0$ being a length  parameter. The coordinates range is: $0{\leqslant}\phi{<}2\pi$, ${-}{\infty}{<}ct/a{<}{+\infty}$, $0{<}\delta{<}r/a{<}{+}\infty$, ${-\infty}{<}z/a{<}{+\infty}$.  
The number $\delta$ determines the radius of a boundary cylindrical surface demarcating the interior material solution from the  exterior vacuum solution. By allowing $\delta{=}0$, the spacetime would become nakedly singular.  
 
The components of the Riemann tensor in the orthonormal tetrad behave as $D_r^{-2}$ with the geodetic radial distance $D_r$ from the symmetry axis. The Kretschmann curvature scalar behaves as $D_r^{-4}$. The curvature decreases slower with $D_r$ than in spacetimes of isolated bodies, for which the corresponding asymptotic behavior is  $D_r^{-3}$ and $D_r^{-6}$, respectively.  
The horizontal planes are not asymptotically Euclidean and close up at infinity, as pointed out by \citet{Bonnor1979}. Namely,  for a circle, the ratio  of the geodetic circumference $2\pi r$ to the geodetic radius $D_r$, being proportional to $(a/r)^{\sfrac{\Qupsilon^2}{(1{+}\Qupsilon)}}$, tends to $0$ as $r{\to}\infty$. 

This particular metric (in a slightly different notation) was presented by \cite{Wilson1920}. He determined the equations of the geodesics and, by comparing with corresponding Newtonian motion, identified a parameter $m{\equiv}\frac{1}{2}\frac{\Qupsilon}{1{+}\Qupsilon}$ (when sufficiently small) as the mass per unit length (in units of $c^2/G$). The Wilson metric would represent a structureless homogeneous and infinitely thin string. It will be shown that the metric is sufficient to study the problem of radial accretion.   

Considering more general Levi-Civita metric would alter in the radial accretion context only the geodetic distance and the relation of the metric parameters to the actual value of the line mass density, while the accretion equations as functions of the radius would remain formally the same. The free parameters in the Levi-Civita metric are determined by matching conditions with the interior solution.
The interior metric could represent a finite-width cylinder of smooth mass distribution characterized by some integrated mass density per unit coordinate height. An interesting account of the matching problem was given by \citet{Bonnor1979} who found a physically reasonable globally regular solution with perfect fluid in the interior. On this example Bonnor confirmed the result of \citet{Marder1958} that
the Levi-Civita metric contains two genuine arbitrary constants determinable
through the matching conditions from the interior solution. Another global solution was presented by \citet{Bonnor1992,Bonnor1999}. 
       
  \subsection{Conditions to be satisfied by the metric form}
              
              The static and cylindrically symmetric vacuum line element can be obtained by imposing several mathematical and physical requirements. One starts with cylindrical coordinates adapted to three Killing vectors $\partial_{t}$, $\partial_{\phi}$  and $\partial_{z}$. For this to be possible in a vacuum region, it is sufficient to assume that the magnitude of one of the Killing vectors, $\partial_t$ or $\partial_{\phi}$, vanishes at one point and apply a theorem 7.1.1 given in \citet{Wald:1984} handbook, from which it also follows that $\partial_{t}$ and $\partial_{\phi}$ will be orthogonal to $\partial_r$ and $\partial_z$. A further condition $g_{\phi t}{=}0$ ensures that the spacetime will be static. Another requirement $g_{\phi z}{=}0$ is a sort of boundary condition that eliminates the gravitation from surface sources distributed at $z{=}{\pm}\infty$ (the $g_{\phi z}$ is a degree of freedom that remains unspecified by Einstein equations and alters the component $g_{rr}$).      
      
      The remaining freedom of  the metric components is used to specify the yet arbitrary radial coordinate $r$.  It is chosen here to be the circumferential radius (analogous to the areal radius in the Schwarzschild spacetime). By construction, such a radius determines the true arc-length on coaxial circular orbits of the Killing field $\partial_{\phi}$. The magnitude of $\partial_{\phi}$ is then $r$ and nonzero on the symmetry axis. 
                                            
If nonempty, he spacetime will not be asymptotically flat, essentially because of the assumed $z$-translation symmetry implying infinite extension of the material source.\footnote{The lack of asymptotical flatness manifests itself in the Newtonian case by the logarithmically divergent potential.} However, the Komar integral \citep{Komar1959} which can be nicely represented as an integral of a surface two-form  involving the timelike Killing vector
    \begin{equation}\label{eq:komarinteg}-\frac{c}{16\pi G}\oiint_{\partial{V_{C}}}\sqrt{-\det{g}}\,\nabla^{\alpha}\xi^{\beta}\,\epsilon_{\alpha\beta\mu\nu}\,\ud{x}^{\mu}{\wedge}\,\ud{x}^{\nu}\end{equation}
    still can be used to define the line mass density (on similar basis as the Gauss surface integral in the case of the Newtonian string). The integration in \eqref{eq:komarinteg} is taken over the surface $\partial{V_C}$ of a solid cylinder segment $V_C$ of some $z$-coordinate height $h$, coaxial with the symmetry axis (effectively, only the side surface of the cylinder contributes to the integral). The amount of mass $m$ enclosed within $V_C$ increases linearly with $h$ and is  independent of the cylinder's diameter,\footnote{By applying the Stokes theorem to the  integral \eqref{eq:komarinteg}, one obtains a volume integral in a vacuum region between coaxial cylindrical segments of the same height. The volume element $3$-form requires a vector, and the only vector constructed  out of second derivatives of the Killing vector $\xi^{\mu}$ and linear in the curvature tensor and $\xi^{\mu}$, is  $R^{\mu}_{\phantom{\mu}\nu}\xi^{\nu}$, which vanishes in the vacuum. Hence, the integral \eqref{eq:komarinteg} must be independent of the cylinder radius.} as it should be for a string with constant line mass density $\mu{=}m/h$. 
     
     There is, however, at least one objection which might be raised to the above definition.
  The metric considered here is not asymptotically flat. Komar integrals are used to define total charges in asymptotically flat exterior vacuum metrics of isolated systems. Asymptotical flatness allows to unambiguously scale Killing fields in CGS units referred to geodetic distances on the sphere at the spatial infinity. In the current context, there is no preferred unique cylindrical surface on which local infinitesimal coordinate and geodetic distances in the directions of all Killing vectors would agree with each other independently of the field strength.      
   
   To remedy this situation, a single universal cylindrical coordinate system $(t,r,\phi,z)$, already scaled with CGS units, can be assumed for all considered metric tensors in the circumferential radius gauge. The components of the tensors are accordingly constrained and, additionally, such that the mass enclosed within coaxial cylindrical segments of unit $z$-coordinate height  is $\upsilon$ by construction (in units of $\frac{c^2}{2G}\,{\sim}6.73{\times}\,10^{27}\mathrm{g}{\cdot}\mathrm{cm}^{-1}$). Then, the Killing vectors attain the canonical form $\partial_t$, $\partial_{\phi}$ and $\partial_z$ (with unit components in the coordinate basis).  Such scaled coordinates coincide with those of the background complete Minkowski spacetime of the weak field regime of extremely small but nonzero line mass density.
   Assuming complete $2\pi$ arc for the angular coordinate in the circumferential radius gauge, effectively eliminates in the flat space limit the residual {\it global deficit angle parameter} admissible under cylindrical symmetry for a cosmic string.  This assumption will allow to regard the accretion process as a continuous relativistic extension of the analogous Newtonian accretion.

The mechanism of cosmic string contribution to the line mass density can be understood by recalling \citet{AStar1963} gravitation theory in $2{+}1$D spacetime, which later became an active field of research starting with \citet{DESER1984220} work. In this theory, vacuum regions must be metrically flat for structural reasons (in three dimensions the vanishing of the Einstein tensor implies that the Riemann tensor is also vanishing). This makes the $2{+}1$D gravitation essentially different from the $3{+}1$D gravitation.  The static space of a single massive particle in the $2{+}1$D model can be obtained by cutting out certain sector of Euclidean plane enclosed by two infinite radial semi-lines with common vertex, and then by connecting the edges of the cutting.  Upon identifying the edges, the straight lines of the original Euclidean plane with the excised wedge describe bent geodesics in the resulting `glued' geometry. Such arising conic curvature singularity in $2{+}1$D effectively acts at a distance as a point mass,  even though there is no gravitational field. The point particle in $2{+}1$D would correspond to a string in $3{+}1$D with the Riemann curvature vanishing. But the effective mass (proportional  to  the deficit angle) is topological in character and does not contribute to the Komar integral (the $g_{tt}$ component of cosmic string is a constant). Such a string would not have a Newtonian limit. 
     
\subsection{\label{sec:lengthscale}The issue of free parameters in the metric form}
     
The requirements stated in the previous section 
are satisfied  by a broad class of vacuum metrics with the following general form involving $3$ pure numbers $A'$ and $B'$ and $Q$ in addition to a length parameter $\alen'$ and a pure number $\upsilon$:     
\begin{eqnarray}\label{eq:metric2}\ud{s}^2{=}-c^2\frac{\upsilon}{Q}A'^2\br{\frac{r}{\alen'}}^{\!2Q}&{}&\!\!\!\!\!\!\!\!(\ud{t})^2+B'^2\!\br{\frac{r}{\alen'}}^{\!\frac{2Q^2}{1{+}Q}}\!(\ud{r})^2 \\
&+&r^2(\ud{\phi})^2+\frac{\upsilon B'^2}{Q A'^2}\!\br{\frac{\alen'}{r}}^{\!\!\frac{2Q}{1{+}Q}}\!\!(\ud{z})^2.\nonumber\end{eqnarray}
To make it simpler to see various symmetries of the metric components, the number of free constants above is higher than actually needed. Einstein equations control the exponents of the radius while allowing for arbitrary constant factors in the metric components. The free factors were constrained such  that the metric leads to a definite value of the  line mass density $\mu$ defined in CGS units by the  integral \eqref{eq:komarinteg}. 
For general diagonal metric, the integration in  \eqref{eq:komarinteg} would give
 $$\mu=\frac{c^2}{2G}{\cdot}\upsilon,\quad \upsilon\equiv
\frac{\sqrt{-g}}{c{\cdot} g_{rr}}\,\partial_{r}\ln{\br{-g_{tt}/c^2}}.$$ 
    Such determined parameter $\upsilon$ will be called the {\it Killing line mass density}.  With metric \eqref{eq:metric2}, the Killing line mass density is identically $\upsilon$ for any $A',B',\alen',Q$ and independent of the cylinder radius. 

It should be noted, that by redefinition of the free constants in \eqref{eq:metric2}, the number $\upsilon$ can be made to appear in other places in the metric and there is no particular meaning to be attached to the current position of $\upsilon$ in \eqref{eq:metric2}. For now it can be assumed that $\alen'{>}0$, $A'{>}0$, $B'{>}0$,  $Q{\neq}{-}1$ and $\upsilon{\cdot}Q{>}0$. Numbers $Q$ and $\upsilon$ can be regarded as independent constants. However, in order to include Minkowski metric as a limit, it should be assumed that $\upsilon{\to}0$  when $Q{\to}0$ in such a way that $\lim_{Q\to0}(\upsilon/Q){\neq}0$.

\medskip

The cylindrical coordinates have been already fixed and given physical meaning, therefore they cannot be rescaled further to absorb free constants in the metric components. Possible spurious degrees of freedom present in these components must be eliminated in another way. The Kretschmann invariant (and all components of the Riemann tensor in the orthonormal basis) expressed through the geodetic distance $D_r$ is a function of only $Q$ (for now, independent of $\upsilon$), for example:
$$R_{abcd}R^{abcd}{=}
\frac{16\,Q^2(1{+}Q)^2}{(1{+}Q{+}Q^2)^3}\frac{1}{D_r^4}.$$
Therefore, the parameter $Q$  is geometrically meaningful.
   
It is worth noting on this occasion, that the $Q$ value can be measured based on a redshift formula. A local redshift measurement  performed at some radius $r$ between cylinders of radii differing by $\Delta r{\ll}r$  gives a number $Z_r$ related to $Q$ through $Q{=}Z_r \frac{r}{\Delta r}$, independently of the path of the light ray between these two cylinders (the required redshift formula will be given later).
   
Furthermore, the metric components have a scaling symmetry $\alen'{\to}\lambda\, \alen'$, $A'{\to} \lambda^{Q}A' $ and $B'{\to}\lambda^{\sfrac{Q^2}{(1{+}Q)}} B'$, leaving the components numerically unaltered. This means that there is a one-parameter family of triples $\alen',A',B'$ defining the same metric. For example, the constant $B'$ can be absorbed by the new length parameter $\alen{=}\lambda\, \alen'$ for appropriate value of $\lambda$, namely, $\lambda{=}(B')^{-\sfrac{(1{+}Q)}{Q^2}}$.  Effectively, one can insert $B'{=}1$ in the metric and keep the other two parameters, which will be denoted $A$ and $\alen$. Hence, without no lost of physical content, the metric components can be reduced to a form 
${{-}g_{tt}}{c^{-2}}{=}\frac{\upsilon A^2}{Q}\br{{r}/{\alen}}^{\!2Q}$,
$g_{rr}{=} \br{{r}/{\alen}}^{\sfrac{2Q^2}{(1{+}Q)}}$, 
$g_{zz}{=}\frac{\upsilon}{Q A^2}\!\br{{\alen}/{r}}^{\sfrac{2Q}{(1{+}Q})}$. The local coordinate and geodesic infinitesimal intervals along the radial direction agree with each other at $r{=}\alen$. Similarly, there is a unique radius $r{=}\blen$ at which the local coordinate and geodesic heights agree with each other along the vertical direction.
  Hence $\blen$ can be used in place of $A$ as they are related through $A^2{\equiv}(\upsilon/Q)(\alen/\blen)^{\sfrac{2Q}{(1{+}Q)}}$.  Finally, the metric tensor can be re-expressed in terms of parameters $\alen$, $\blen$, $Q$ and $\upsilon$ with definite geometric and physical meaning: 
    \begin{eqnarray}\label{eq:metricab}\!\!\!\ud{s}^2{=}
    -c^2\frac{\upsilon^2}{Q^2}\br{\frac{\blen}{\alen}}^{\!\!\frac{2Q^2}{1{+}Q}}\!\br{\frac{r}{\blen}}^{\!2Q}\!(\ud{t})^2&& \\ +\br{\frac{r}{\alen}}^{\!\frac{2Q^2}{1{+}Q}}\!(\ud{r})^2  \nonumber
  +r^2(\ud{\phi})^2&+&\br{\!\frac{\blen}{r}\!}^{\!\!\frac{2Q}{1{+}Q}}\!\!(\ud{z})^2.
   \end{eqnarray}
   For the metric to become almost Minkowskian, it is necessary that $Q$ be extremely small.  
    But  in the weak field  limit also $\upsilon$ must be extremely small, and so the two constants $Q$ and $\upsilon$ cannot be independent.  
       
     The leading term of a power series expansion in $Q$ of the metric component $g_{tt}$  in \eqref{eq:metricab} reads $g_{tt}{\sim}{-}c^2(\upsilon/Q)^2\br{1{+}2Q\ln{(r/\blen)}{+}o(Q)}$ and it must be finite when $Q{\to}0$. Comparison of $\partial_rg_{tt}$ with the force exerted by the Newtonian string characterized by the line mass density $\frac{c^2}{2G}{\cdot}\upsilon$, shows that  $Q{=}\upsilon{+}o(\upsilon)$, and hence $g_{tt}{\sim}{-}c^2\br{1{+}2\upsilon\ln{(r/\blen)}{+}o(1)}$ in this limit. But then $A^2{\to}1$, consistently with the requirement that $t$ is identified in the weak field limit with the Newtonian time scaled in CGS units. The ratio $\alen/\blen$ comes into play in the postnewtonian, second order expansion in $Q$.  
      
      Although  $Q{\sim}\upsilon$ in the weak field regime, the two numbers might be different in the strong field regime. 
            For stronger fields, $Q$ will be related to $\upsilon$ through some function $f$ of metric parameters. But this function, as well as its arguments, must be dimensionless, hence  
 $$\upsilon{=}Q f(\alen/\blen,Q),
   \quad\lim\limits_{Q{\to}0}f(\alen/\blen,Q){=}1.$$  In particular, all three Killing vectors might be commensurate,\footnote{
   The Killing vectors will be called {\it commensurate} when they are normalized in the same reference hyper-surface of Killing symmetries on which the coordinate lengths in the coordinates adapted to these symmetries measure geodetic distances. This notion derives from normalization of Killing vectors at spatial infinity in asymptotically flat spacetimes (when the vectors are commensurate on the sphere at infinity). In cylindrical coordinates it is then required that $|\partial_t|{=}c^2$, $|\partial_z|{=}1$ and $|\partial_{\phi}|{=}r^2$ on a unique cylindrical surface. For the metric \eqref{eq:metricab} this is possible only at $r{=}\blen$, then \eqref{eq:commensr} must be satisfied.} in which case $f{=}(\alen/\blen)^{\sfrac{Q^2}{(1{+}Q)}}$, and so
   \begin{equation}\label{eq:commensr}\upsilon\equiv Q\br{{\alen}/{\blen}}^{\!\frac{Q^2}{1{+}Q}}.\end{equation} It is not a priori obvious that Killing vectors should be commensurate. However, one can observe  that general Levi-Civita metrics (usually expressed in the literature in Weyl coordinates) satisfy this  condition -- for example, this is the case for the general metric considered by \citet{Bonnor1999}.\footnote{The transformation to the circumferential radius gauge is: $r{=}L P (R/L)^{1{-}2m}$, $\alen{=}LP((1{-}2m)P)^{\sfrac{(1{-}2m)}{(4m^2)}}$,  $\blen{=}L P$,  $Q{=}\sfrac{2m}{(1{-}2m)}$,  where $R$ is the radius in Weyl coordinates, $L$ a unit of length, $P$ and $m$ is a pair of free parameters defining the general Levi-Civita metric as given in equation 2 in \citep{Bonnor1999}.  The quantity $Q\br{\alen/\blen}^{\sfrac{Q^2}{(1{+}Q})}$ for such expressed $\alen$ and $\blen$ evaluates to $2Pm$, which overlaps with the line mass  density obtained with the help of Komar integral \eqref{eq:komarinteg} for the same metric. This shows that the Killing vectors  indeed are commensurate.} Nevertheless, the results of the next section are obtained without assuming commensurability. 
   
   With certainty, the particular form of $f$ will be dependent on the interior metric which the exterior vacuum metric is to be matched to at some boundary cylinder. Examples of matching of the vacuum solution to physically viable interiors were given by \citet{Marder1958}, \citet{Bonnor1979, Bonnor1992, Bonnor1999}.
 
\medskip         
\subsection{\label{sec:arole}Physical observables as functions of the ratio $\alen/\blen$}

Below, considered is a class of metric tensors \eqref{eq:metricab}  described by the same parameter $Q$ and various $\alen/\blen$  in the same coordinate system $(t,r,\phi,z)$ already scaled in CGS units.\footnote{Effectively, only the length $\alen$ will appear as a parameter in the considered equations through a ratio $r/\alen$, however the dependence of an observable described by the equation on the parameter $\alen/\blen$ is evident since both $r$ and $\alen$ can be scaled in units of $\blen$ considered here as a fixed parameter.} The local physical observables will be referred to the orthonormal basis of a locally static observer
$$\begin{array}{ll}
e_t=c^{-1}\!\br{Q/\upsilon} (\alen/\blen)^{\frac{Q^2}{1{+}Q}}(\blen/r)^{Q}\partial_t, 
&  e_{\phi}=r^{-1}\partial_{\phi}, \\
e_z=(r/\blen)^{\frac{Q}{1{+}Q}}\partial_z, 
& e_r=(\alen/r)^{\frac{Q^2}{1{+}Q}}\partial_r.
\end{array}$$ 
A simple example of a geometric  observable is provided by a combination of the geodesic distance $D_r$ and the circumferential radius $r$. The ratio $D_r/r$  changes with $\alen/\blen$ in a nonlinear manner: 
$D_r/r{=}\frac{1}{1{+}q}(r/\alen)^q$, $q{=}\sfrac{Q^2}{(1{+}Q)}$. At the same time, the ratio of such  ratios taken at two different radii, which reads 
 $(r_1/r_2)^{q}$, is independent of  $\alen/\blen$. Similarly, physical observables considered as functions of $r$ will be dependent or independent of $\alen/\blen$.

\medskip

\noindent$\bullet$ {\it The frequency shift formula}. Let $k{=}(\omega/c){\cdot}(e_t{+}n)$ be a  wave vector of a photon moving in some arbitrary direction described locally by a unit vector $n$ orthogonal to $e_t$. The frequency $\omega$ is defined as that perceived  by local static observers  $\omega{\equiv}{-}c\,k.e_t$. The scalar $k.\partial_t{=}{-}(\upsilon/Q)(\blen/\alen)^{\sfrac{Q^2}{(1{+}Q)}}(r/\blen)^{Q}{\cdot}\omega$ involving the Killing vector $\partial_t$  is conserved along geodesics. Hence, the frequencies $\omega_i$ recorded at different radii $r_i$ do not depend on other coordinates and satisfy the power law $${\omega_2}/{\omega_1}{=}\br{r_1/r_2}^{Q}.$$  Through the resulting formula $Q{=}{-}\frac{r}{\omega}\frac{\ud{\omega}}{\ud{r}}$, the law links the observable $Q$ to the redshift measurements, as discussed at the end of \secref{sec:lengthscale}. It is noteworthy that the frequency ratio is scale invariant -- it depends only on the ratio of the circumferential radii.

\medskip

\noindent $\bullet$ {\it The Keppler problem.}  Since the momentum is conserved along the axial symmetry axis,  it is enough to consider only the motion in a horizontal plane of constant $z$.  The trajectory equation can be derived by making use of conserved quantities and then succinctly represented in a form 
involving only quantities with clear physical interpretation (the derivation is skipped here). Namely, the trajectory 
of free fall in the polar parametrization $r{=}r(\phi)$ is described by the equation
\begin{equation}\label{eq:traj}\frac{1}{r}\frac{\ud{r}}{\ud{\phi}}{=}\br{\frac{\alen}{r}}^{\!\!\frac{Q^2}{1{+}Q}}\tan{\gamma}.\end{equation}
Here, $\gamma$ (given in full form below) is the true angle  formed in the curved geometry of the horizontal plane and directed from the versor $e_{\phi}$ of the local azimuthal direction to the local versor tangent to the trajectory. Except for integer multiples of $\pi/2$, $\gamma$ differs from the corresponding Euclidean angle in the horizontal plane regarded as a flat plane (with the same polar coordinates).
Given the initial angle $\dfty$ and velocity $\bfty$ specified at some radius $\rfty$ on the trajectory, the corresponding values at any other radius on this trajectory can be found from the equations 
$$\cos{\gamma}{=}\frac{\bfty}{\beta}\br{\!\frac{\rfty}{r}\!}^{\!1{-}Q}\!\cos{\alpfty},\quad \beta{=}\sqrt{1{-}(r/\rfty)^{2Q}(1{-}\bfty^2)}.$$
The equations allow to express $\tan{\gamma}$ in the trajectory equation \eqref{eq:traj} as a function of only $r$. The scalar $\beta$ is the magnitude of total physical velocity (in units of the speed of light) as measured with respect to locally static observers, that is, $\beta$ is numerically equal to the hyperbolic tangent of the hyperbolic angle between the respective fourvelocities.

The length $\alen$ can be eliminated from \eqref{eq:traj} by considering an auxiliary variable  $\phi'$: $\phi(\phi'){\equiv}(\rfty/\alen)^{\sfrac{Q^2}{(1{+}Q)}}\phi'$. Suppose that a solution $\rho(\phi'){\equiv}r(\phi(\phi'))/\rfty$ of the transformed equation has been found. This reference solution (corresponding to $\alen{=}\rfty$) can be represented  as a unique curve $r(\phi){=}\rfty \rho(\phi)$ in the same polar coordinates.  Then the solutions $r_{\alen}(\phi){=}\rfty \rho(\phi'(\phi))$ of the original equation corresponding to $\alen$-dependent metric tensors  can be obtained and represented on the same Euclidean plane by rotating points lying on the reference curve through an angle $\Delta\phi{=}\phi'{\cdot}((\rfty/\alen)^{\sfrac{Q^2}{(1{+}Q)}}{-}1)$,  remembering that $\phi$ is identified with $\phi{+}2\pi$ (the rotation angle will be large since the characteristic width of the gravitation source, being comparable to $\alen$,  will usually be much smaller than $\rfty$).

For such curves, representing orbits passing through a common initial point $(\phi,r){=}(0,\rfty)$ with the same initial data $\bfty$ and $\alpfty$, the initial Euclidean angle changes with $\alen$.
As the angular parameter $\phi$ increases, the corresponding points on the curves move away from each other along the radial direction. However, at the intersection points of the curves with a circle centered at the origin, both the velocities $\beta$ and the angles $\gamma$ are the same, because they are functions of $r/\rfty$ only. All these orbits are confined between two unique bounding circles which are the loci of turning points of the orbits. The radii $r_i{<}\rfty$ and $r_e{>}\rfty$ of the bounding circles are independent of $\alen$ and they are determined by two roots of the equation for $\cos{\gamma}$ defined above, with $0$ substituted for $\gamma$. The shape of a given orbit is governed by the ratio $\rfty/\alen$. The amount of angle a point on the orbit travels about the origin between consecutive turning points grows with decreasing $\alen$ and so grows with the increasing geodetic distance $D_{\infty}{=}\rfty\frac{1{+}Q}{1{+}Q{+}Q^2}(\rfty/\alen)^{\sfrac{Q^2}{(1{+}Q)}}$. This result agrees with the intuition but it also can be interpreted in terms of deficit or excess angle governed by the ratio $\alen/\blen$. The latter interpretation is suggested by the previous construction of the substitute angle $\phi'$.

It is important to note that the quantities  $\Delta\phi(r)$ and $D_{\infty}/\rfty{-}1$  are of second order in $Q$, that is, corresponding to corrections of order $G^2{/}c^4$. Thus, the ratio $\alen/\blen$ is of no importance in the Newtonian limit of the motion.
Indeed, in the non-relativistic regime, the numbers $Q{\sim}\upsilon$ must be extremely small, as described in \secref{sec:lengthscale}. Then, the background metric in the horizontal plane is flat and the free fall occurs in an effective logarithmic potential in which any length scale can be absorbed by the arbitrary constant in the gravitational potential and thus has no physical meaning.

For a purely radial free fall  ($\gamma{\equiv}\pi/2$) the functional dependence between the velocity and the radius is the same as for the spiraling motion just considered. Consequently, the velocities at different radii are related by a function of only the ratio of the radii.
A similar conclusion will be true also for the radial accretion process investigated later.
       
\subsection{Wilson string. A limitation on the line mass density}
 
The metric \eqref{eq:metricab} is described by effectively two numbers:  a Komar line mass density $\upsilon$ and a ratio of two lengths $\alen/\blen$. Through the matching conditions at the boundary cylinder, the numbers are related to the geometric parameter $Q$ and to some two length scales characterizing the interior medium. A single length scale must always be present on dimensional grounds, to allow for non-integer powers of the radius. If there are two length scales present, then they are connected with length parameters $\alen$ and $\blen$. The ratio $\alen/\blen$ is then critical. Although it is irrelevant in the Newtonian context,  its role is important in the ultra-relativistic regime, as discussed in \secref{sec:arole}.  The presence of two length scales  could be an artifact of a characteristic width of the interior medium or  unsymmetrical tension within it.
   
It can be hypothesized  that $\alen{=}\blen$ in the limit of infinitesimally thin homogeneous string. Assuming commensurability of Killing vectors, this gives $\upsilon{=}Q$, and so the metric \eqref{eq:metricab} reduces to the particular Wilson form \eqref{eq:metric}. Then, the number $Q$ has a direct interpretation in terms of the line mass density (in units of $\tfrac{c^2}{2G}$). Since $\upsilon{>}0$ in the Newtonian limit, one should postulate that $\upsilon$ is non-negative in general. Furthermore, the velocity in free motion on circular orbits relative to local stationary (Killing) observers is $c\sqrt{\upsilon}$, independently of the radius. Such geodesics can be timelike only for $\upsilon{<}1$. This provides an upper bound for $\upsilon$ if such orbits are to be always possible. There is also an independent indication due to \citet{Lathrop_1980} that $\upsilon{=}1$ might be the upper limit for general infinite cylinders. It is henceforth assumed for Wilson metric that $$0{<}\upsilon{<}1$$ (for the metric \eqref{eq:metricab} the circular velocity would be $c\sqrt{Q}$, and so the analogous limitation is $0{<}Q{<}1$).
          
It is clear from the considerations, that for any $\upsilon$ in the assumed range, the quantity $\mu{\equiv}\,\upsilon{\cdot}\frac{c^2}{2G}$ has a consistent physical interpretation as the mass per unit coordinate length uniformly distributed along the cylinder's axis. This motivated to parametrize the exponents in the metric tensor just in terms of $\upsilon$ as in \eqref{eq:metric}. 

\section{\label{sec:taub}The model of accreting matter}

\subsection{Planar motion of perfect fluid in Levi-Civita spacetime}  The stress tensor of perfect fluid is described in its rest frame by a total energy density $e$, a particle density  $n$, and a pressure $p$:
$\Tud{T}{\mu}{\nu}{=}(e{+}p)u^{\mu}u_{\nu}{+}p\,
\Tud{\delta}{\mu}{\nu}$.\footnote{For example, the components  of $T$ for purely radial inward flow four-velocity $u{=}c{\cdot}\frac{e_t -\beta e_r}{\sqrt{1-\beta^2}}$ as given in the basis of static observers (with four-velocity $c{\cdot}e_t$ and spatial versors $e_r$, $e_{\phi}$, $e_z$  defined earlier for the metric \eqref{eq:metricab}) read
\begin{equation*}\left[\begin{array}{cccc}
\frac{e+p\beta^2}{1-\beta^2} & \frac{(e+p)\beta}{1-\beta^2} & 0 & 0\\
\frac{(e+p)\beta}{1-\beta^2} & \frac{e\beta^2+p}{1-\beta^2} & 0 & 0\\
0 & 0 & p & 0 \\
0 & 0 & 0 & p \end{array}\right].\end{equation*}}
The hydrodynamical equations consist of the  conservation $\nabla_{\mu}\Tud{T}{\mu}{\nu}{=}0$ and the continuity $\nabla_{\mu}(nu^{\mu}){=}0$ constraints. Projected onto the flow four-velocity vector $u^{\nu}$, the equations imply the following condition \begin{equation}\label{eq:isoentr} \ud{\br{\frac{e+p}{n}}}=\frac{1}{n}\,\ud{p}.\end{equation}
This result is equivalent to the first law of thermodynamics for isentropic process with a conserved number of particles. Supplemented with the equations $\nabla_{\mu}\xi_{\nu}{+}\nabla_{\nu}\xi_{\mu}{=}0$ satisfied by Killing vectors $\xi{=}\partial_{t}$ or $\xi{=}\partial_{\phi}$, the constraints also imply
conservation laws $\nabla_{\mu}j^{\mu}{=}0$ for the currents $j^{\mu}{=}\Tud{T}{\mu}{\nu}\xi^{\nu}$. The corresponding first integrals reduce to $n^{-1}(e{+}p)(u\xi){\propto}\epsilon$ and $n^{-1}(e{+}p)(u\eta){\propto}\lambda$, where it has been assumed that the space-time symmetries concern also the flow -- in particular, $\xi^{\mu}\partial_{\mu}p{=}0$. In the case of planar flow the specific enthalpy $\epsilon$ and the specific angular momentum $\lambda$ constants (varying from worldline to worldline) read    
\begin{equation}\label{eq:1stinteg1and2}
\!\;\!\:\frac{(e{+}p){\cdot}(r/\alen)^{Q}}{mc^2\,n\sqrt{1{-}\beta_r^2{-}\beta_{\phi}^2}}
\!=\!\epsilon,\quad \\ \quad
\frac{(e{+}p)\beta_{\phi}{\cdot}(r/\alen)}{mc^2n\sqrt{1{-}\beta_r^2{-}\beta_{\phi}^2}}
\!=\!\lambda
\end{equation}
(both constants as defined here are dimensionless).
Here, $\beta_r$ and $\beta_{\phi}$ are the physical radial and azimuthal components of the velocity flow measured by the static observer: $\beta_i{=}(\frac{u}{u.e_t}{+}e_t).e_i$, where $e_i$ are unit vectors in the radial and azimuthal direction, $u.e_{z}{=}0$, and $e_t$ is a unit vector directed along $\partial_{t}$. The form of the first equation in \eqref{eq:1stinteg1and2} implies 
$$\epsilon>0.$$
The third conserved quantity, the accretion rate $\kappa$, follows from the integration of the  continuity constraint (simplifying here to $\partial_r\!\br{nu^r\sqrt{-g}}{=}0$), hence 
    \begin{equation}\label{eq:1stinteg3}\frac{(r/\alen)^{Q{+}\frac{1}{1{+}Q}}\,(n/n_o)\,\beta_r}{\sqrt{1{-}\beta_r^2{-}\beta_{\phi}^2}}=-\mathcal{\kappa}.\end{equation}
    A reference unit of concentration $n_o$ has been introduced to make $\kappa$ dimensionless (for the accretion $\kappa{>}0$ and $\beta_r{<}0$ is assumed).

\subsection{The equation of state}

The considerations so far were valid for general perfect fluid under the assumed symmetries. Let the set of equations be supplemented with classical ideal gas equation of state \cite{PhysRevD.99.084035} enumerated with the polytropic ($\chi$) or the Poisson adiabatic ($\alpha{=}1{+}\chi^{-1}$) index, assumed to be constant in the range of temperatures considered: $p{=}n k_B T$ and $e{=}n\,(m\,c^2 {+}\chi k_B\, T)$. The isentropic condition \eqref{eq:isoentr} is solved by the following $\ydens$-parameterization (here, $n_o{\equiv}n(T_o)$ with $T_o{\equiv}\frac{mc^2}{k_B}$)
$$\frac{n}{n_o}\!=\!\ydens,\quad\! 
\frac{p}{n_o m c^2}\!=\!{\ydens}^{\alpha},\quad\!
\frac{e}{n_o m c^2}\!=\!{\ydens}\!+\!\frac{1}{\alpha\!-\!1}\,{\ydens}^{\alpha},\quad\! \alpha{>}1$$
or, defining $\theta\!\equiv\!\frac{k_B\, T}{m\,c^2}={\ydens}^{\alpha{-}1}$, by the $\theta$-parametrization:
$$\frac{n}{n_o}\!=\!\theta^{\chi},\quad 
\frac{p}{n_o m c^2}\!=\!\theta^{\chi\!+\!1},\quad
\frac{e}{n_o m c^2}\!=\!\theta^{\chi}\!+\!\chi\,\theta^{\chi\!+\!1}.$$
The sonic velocity $c_s$ (in units of $c$) is
\begin{equation}\label{eq:sonicspeed1}\!
c_s^2{\equiv} \frac{\ud{p}}{\ud{e}} {=}\frac{\alpha \,
    {\ydens}^{\alpha-1}}
    { 1{+}
     \tfrac{\alpha}{\alpha{-}1} \,
       {\ydens}^{\alpha-1}
       }{=}
       \frac{(1{+}\chi)\theta}{\chi(1{+}(1{+}\chi)\theta)}{<}\frac{1}{\chi}{=}\alpha{-}1. \end{equation}       
The $\alpha$ is usually constrained by \cite{PhysRev.74.328} limit $\alpha{\leqslant}2$ from the subluminal sonic speed requirement, although one can consistently assume also $\alpha{>}2$ in some cases \citep{PhysRevD.99.084035}. The former limitation would be applicable to the model considered here. As it turns out later, for $\alpha{>}2$ the integral curves on the $r$-$c_s$ plane enter the region with superluminal speed  close to the symmetry axis, while the speed of accreting matter remains subluminal.

The above parametrization constrains the gas pressure $p$ and the gas `rest mass' energy density $\varrho{\equiv} mc^2n$ by a power law $p\!\;{\propto}\varrho^{\alpha}$ characteristic of polytropic matter. Described by simple thermodynamic relations, this form of matter sometimes allows a complete solution of the gas dynamics \cite{Landau1987Fluid}. Instead of assuming the ideal gas as above, the same power law could have been assumed as a starting point, supplemented with the same relativistic expression for the total energy density involving the `rest mass' energy density and the thermal energy density, as before. This kind of matter would be isentropic only if its temperature were defined as $\theta{\equiv}p/\varrho$ (this can be verified by solving the differential equation  \eqref{eq:isoentr} with the condition $p{\to}0$ as $\theta{\to}0$). 

Polytropic matter with such defined temperature was assumed in the classical paper on relativistic accretion \citep{1972Ap&SS..15..153M}. Therein, an auxiliary variable $V^2{\equiv}\frac{\ud{\,\ln(e+p)}}{\ud{\,\ln{\varrho}}}{-}1$ was used. It is the same function of the temperature as $c_s^2$  in  \eqref{eq:sonicspeed1}. This coincidence can be explained by making a general observation that for isentropic flow satisfying \eqref{eq:isoentr}, the $V^2$ with the logarithmic derivatives, reduces (independently of the equation of state) to the definition of the sonic velocity squared, that is,  $\frac{\ud{\,\ln(e+p)}}{\ud{\,\ln{\varrho}}}-1{=}\frac{dp}{de}$. In conjunction with the previous observation that the conservation laws of perfect fluid imply isentropic flow, one comes to the conclusion that \citet{1972Ap&SS..15..153M} effectively considers a kind of matter equivalent to ideal gas with stress tensor of perfect fluid. 

For this reason the accretion model investigated here may be regarded as a cylindrical analog of the \citet{1972Ap&SS..15..153M} spherical model.
Having noticed this, a remark similar to that given therein can be made that there might be more appropriate equations of state to be considered, however, for more sophisticated equations of state the analysis would become computationally more difficult and dominated by technicalities.

\subsection{The non-relativistic accretion limit}
The first integrals \eqref{eq:1stinteg1and2} and \eqref{eq:1stinteg3} provide three constraints on four variables $r,{\ydens},\beta_r,\beta_\phi$. If one chooses to eliminate the velocities, then ${\ydens}$ can be regarded as a function of $r$. In this case there is a linear combination of integrals $\epsilon^2,\lambda^2,\kappa^2$ with coefficients chosen such that the combination sums up to unity. On rearranging terms one arrives at a Hamiltonian-like constraint involving three constants $\epsilon, \lambda$ and $\kappa$ (as discussed earlier, for non-relativistic limit it is sufficient to consider the metric \eqref{eq:metric} as a starting point):
\begin{equation}\label{eq:hamiltonian}\frac{{\epsilon }^2}{x^{2\upsilon }} {= }
  \frac{{\lambda }^2}{x^2} {+} 
   \left( 1 +  \tfrac{\alpha}{\alpha{-}1}
         {\ydens}^{\alpha{-}1} \right)^{\!\!2}\!
    \left( 1 {+} \frac{{\kappa }^2}{x^2{\ydens}^2}\,
       {x^{-\frac{2\upsilon^2}{1 {+} \upsilon }}} \right)\!,
       \end{equation} where $x{=}r/a$. Solving for ${\ydens}$ as a function of $x$ and noticing that $$\beta_{\phi}=\frac{x^{\upsilon}}{\epsilon}\frac{\lambda}{x},$$ the $\beta_r$ component can be determined  as a function of $x$ from any of the previous first integrals. 
       
In the non-relativistic limit, $\upsilon{\to}0$ and $e{+}p{\,\sim\,} n mc^2$, hence $x^{\upsilon}{\sim} 1{+}\upsilon \ln{x}$, and so $\epsilon^2{\sim}1{+}2\tilde{\epsilon}$. Expanding \eqref{eq:hamiltonian} to the linear order in  $\tilde{\epsilon}$ and $\upsilon$ (both proportional to $c^{{-}2}$), gives
       $$\tilde{\epsilon}=\frac{1}{2}\frac{\lambda^2}{x^2}+\frac{1}{2}\frac{\kappa^2}{x^2{\ydens}^2}+\tfrac{\alpha}{\alpha-1}\,{\ydens}^{\alpha-1}+\upsilon \ln{x}.$$ This expression overlaps (modulo units) with the Hamiltonian describing spiraling accretion problem in the logarithmic Newtonian potential  investigated in \citep{Bratek_2019} which is fully solvable in terms of the Lambert ${W}$ function:
\newcommand{\kyh}{\frac{\eexp{-2\psi}}{\upsilon}\!\br{\!\lambda^2{+}\frac{\kappa^2}{{\ydens}^2}}}
\newcommand{\eyh}{\!\tilde{\epsilon\,}{-}\frac{{\alpha}\!\:{\ydens}^{\alpha{-}1}}{\alpha{-}1}\!}
{{\small $$
x({\ydens}){=}\exp{\!\sq{\psi{+}\frac{1}{2}\,W\!\!\br{\!\!-\kyh\!\!}\!}},
\ \psi{=}\frac{1}{\upsilon}\!\br{\eyh}
$$ }}
\!\!\!(the complete level line employs both branches of the double valued function $W$). The Lambert ${W}$ function appeared earlier in many contexts of radial spherical flows \cite{cranmer2004, Ciotti_2017, Ciotti_2018}.
       
\section{\label{sec:qualitres}Qualitative results for purely radial accretion}
         
In the purely radial accretion process one is concerned only with functional dependence of mechanical and thermodynamical quantities between two states at two different radii. Since such a comparison is made through conserved quantities \eqref{eq:1stinteg1and2} and \eqref{eq:1stinteg3}, it immediately follows that the functions will involve only ratios of the radii, while any length parameter specifying the metric tensor, such as $\alen$ or $\blen$, gets canceled out.\footnote{Throughout this section a dimensionless radial variable $x{\equiv}{r/a}$ is used.}
  Moreover, the length $\alen$ enters the integrals \eqref{eq:1stinteg1and2} and \eqref{eq:1stinteg3} through powers of the ratio $r/\alen$. Hence one can introduce any other, more convenient reference radius in place of $\alen$, and this will only result in some redefinition of the constants of motion. In this respect, the only difference between using metric \eqref{eq:metric} or \eqref{eq:metricab} is the interpretation of the number $Q$ in terms of $\upsilon$ and $\alen/\blen$, while the solutions will be formally the same. Therefore, from now on it will be assumed that $Q{=}\upsilon$ and $\alen{=}\blen$, and so  the results for Wilson metric form \eqref{eq:metric} will be given (however, to obtain results for the general Levi-Civita metric \eqref{eq:metricab}, one has to substitute $Q$ in place of $\upsilon$ in all equations below). The analysis of the flow will be considered for 
 $\alpha$ and $\upsilon$ regarded as free parameters and some results will be illustrated for particular values of these parameters.
 
In what follows  the centrifugal force term is disregarded ($\lambda{=}0$). 
The first integrals \eqref{eq:1stinteg1and2} and \eqref{eq:1stinteg3} reduce to 
$\Epsilon(x,{\ydens},\beta){=}\epsilon$ and $\Kappa(x,{\ydens},\beta){=}\kappa$, where  
\begin{equation*}\Epsilon(x,{\ydens},\beta){\equiv}\frac{1{+}\frac{\alpha}{\alpha{-}1}
{\ydens}^{\alpha{-}1}}{\sqrt{1{-}\beta^2}}\,x^{\upsilon},\quad   \Kappa(x,{\ydens},\beta){\equiv}\,
\frac{\beta\,{\ydens}\,x^{\frac{\upsilon}{\au{-}1}}}{\sqrt{1{-}\beta^2}}.
\end{equation*}
\!\!Here, $x{=}r/a$,  $\beta{\equiv}{-\beta_r}$, and  $\au$ is a {\it critical adiabatic index} (its criticality will become clear  later) defined by \begin{equation}\label{eq:au}\au {\equiv} 1{+}\frac{\upsilon\left( 1{ +} \upsilon  \right) }
  {1 {+} \upsilon  {+} {\upsilon }^2}<\frac{5}{3}\qquad ({\rm since}\,\, \upsilon{<}1).\end{equation} 
The integral $\Epsilon$ expresses the conservation of specific enthalpy along streamlines and $\Kappa$ expresses the continuity of the flow. If $\kappa$ is kept constant, then $\epsilon$ can be regarded as a function. Similarly, for constant $\epsilon$, $\kappa$ can be regarded as another function. Both functions represent some surfaces over a two-dimensional plane. Particular solutions are obtained when $\epsilon$ and $\kappa$ are both fixed. The solutions can be represented locally as level lines of the surfaces. On these surfaces, critical points are defined as those at which the gradient vanishes. From the astrophysical standpoint, interesting are critical values of the accretion. It is thus appropriate to investigate just the {\it accretion rate surface} to see qualitative features of solutions. 
           
The enthalpy per particle is set by a parameter $\epsilon$ which is arbitrary, however it can be assumed that its value is defined by a representative of initial states enumerated with various accretion rates. On making the following redefinitions
$$x\to\,\,u\equiv x^{\upsilon}{/}\epsilon,\qquad \kappa\to{\,\,\varkappa}\equiv\frac{\kappa{/}\epsilon^{\frac{1}{\au{-}1}}}{\br{\!1{-}1{/}\alpha}^{\frac{1}{\alpha{-}1}}},$$
the first integrals can now be recast to a form not involving $\epsilon$ explicitly
\begin{equation}\label{eq:radial_integrals}\frac{1{+}\frac{\alpha}{\alpha{-}1}{\ydens}^{\alpha{-}1}}{\sqrt{1{-}\beta^2}}{=}\frac{1}{u}
\ \ \mathrm{and}\ \ \frac{\beta\,{\ydens}\,u^{\frac{1}{\au{-}1}}}{\br{\!1{-}1{/}\alpha}^{\frac{1}{\alpha{-}1}}\!\sqrt{1{-}\beta^2}}{=}\varkappa.\end{equation}
(the form of the first equation shows that $u$ cannot exceed unity, and so $x^{\upsilon}$ is bounded from above by $\epsilon$). The free parameters remaining in these equations define the qualitative structure of the flow. The constant $\alpha$ enumerates the polytropic matter forms. 
The constant $\au$ (or $\upsilon$) enumerates the background gravitational field strength.
The constant $\varkappa$  is dynamical in nature. It enumerates integral curves with various intensity of the flow. Therefore, it can be regarded  as a function of physical state akin to a Hamiltonian defined over a phase space. This idea will be employed in what follows.

Upon eliminating ${\ydens}$ from \eqref{eq:radial_integrals}, the solutions will be investigated as level lines of an {\it accretion rate surface} $A(u,\beta)$ defined over a phase space with $u$ and $\beta$ regarded as independent coordinates:\footnote{The coordinates $x$ and $\beta$ were chosen because they are naturally bounded. Then the level lines of the accretion surface can be represented over a compact domain convenient for phase diagrams.}
 \begin{equation}\label{eq:accrsurface}
 A(u,\beta){\equiv}\frac{\beta{u}^{\frac{1}{\au{-}1}}}{\sqrt{1{-}\beta^2}}\!\sq{\!\frac{\sqrt{1{-}\beta^2}}{u}{\,-}1\! 
        }^
    {\!\!\frac{1}{\alpha-1}}\!\!\!,\quad\!\!\! u{<}\sqrt{1{-}\beta^2}.\!\!\end{equation} 
In principle, the level lines could be given in an exact form in terms of a new special function obtained according to the Lagrange theory of finding solutions of equations of the form $Y{=}a{+}X\,\Phi(Y)$. This could be done in a manner analogous to introducing the Lambert W-function as a series expansion solution to the equation $Y{-}Xe^{-Y}{=}0$. Specifically, with $Y{\equiv}{u}/{\sqrt{1{-}\beta^2}}$ and $X{\equiv}({\varkappa}/{\beta})^{\alpha{-}1}{\!/\!}\br{1{-}\beta^2
                }^{\!\frac{(\alpha{-}1)(2-\au)}{2(\au{-}1)}}$, one obtains an example of Lambert transcendental equation $Y{+}X\, Y^{\frac{\au{-}\alpha}{\au-1}}{=}1$. The resulting $\Phi(Y)$ is analytic at $Y{=}1$.  Then one could deduce that the solution is a function $Y(\varkappa^{\alpha{-}1})$ (dependent on parameters $u,\beta$) which for $\varkappa$ small enough is analytic at $\varkappa{=}0$ and attaining a value $Y{=}1$ at that point. Then the series expansion could be constructed term by term (an  interesting account of an approach for solving a broad class of Lambert transcendental equations is presented by \citet{perovich2011}).
                
In what follows, the solutions will be investigated  in a qualitative manner akin to that presented in \citep{Bratek_2019}, only later some solutions will be needed in approximate analytical form.
                
    \subsection{\label{sec:shocks}The principal and conjugate critical lines}
    
A turning point can be identified on an integral line defined by the constraints $\Epsilon(x,{\ydens},\beta){=}\epsilon$ and $\Kappa(x,{\ydens},\beta){=}\kappa$ by imposing the additional condition \begin{equation}\label{eq:shock_line_def}
\partial_{\ydens}\Kappa\partial_{\beta}\Epsilon{-}\partial_{\beta}\Kappa\partial_{\ydens}\Epsilon{=}0.
\end{equation} The reason behind  \eqref{eq:shock_line_def} is the requirement that a first order variation $\delta\Kappa(x,{\ydens},\beta)$  should vanish at a point  for a vanishing variation $\delta x$
 whenever $x,{\ydens},\beta$ are constrained by the condition $\Epsilon(x,{\ydens},\beta){=}\epsilon$. The condition \eqref{eq:shock_line_def} is necessary for a point to be a turning point simultaneously on both $x$-${\ydens}$ and $x$-$\beta$ planes.  The total differentials of the  constraints $\Epsilon(x,{\ydens},\beta){=}\epsilon$ and $\Kappa(x,{\ydens},\beta){=}\kappa$ must be vanishing on a given level line. Then, for $\ud{x}{=}0$ the differentials $\ud{\beta}$ and $\ud{\ydens}$ would also be vanishing only when  the condition \eqref{eq:shock_line_def} were violated. If, on the contrary, \eqref{eq:shock_line_def}  is satisfied at some point, then the two differentials could remain nonzero for $\ud{x}{=}0$,  and so the derivatives $\beta'(x)$ and ${\ydens}'(x)$ would diverge at that point.

When $\Kappa$ changes as a function of three variables constrained  by two conditions, $\Epsilon(x,{\ydens},\beta){=}\epsilon$ and \eqref{eq:shock_line_def}, the turning point moves across the level lines of the accretion surface, drawing an intersection line through all level lines. 
The condition \eqref{eq:shock_line_def} is form-invariant with respect to replacing integrals $\Kappa$ and $\Epsilon$ with any other two independent combinations of them. It is also form-invariant with respect to coordinate transformations that mix coordinates of state $\beta$ and ${\ydens}$. Thus, the intersection line has an absolute, coordinate independent meaning, and so it will be referred to as the {\it principal critical line}.
Using the definitions of $\Epsilon$ and $\Kappa$, the condition \eqref{eq:shock_line_def} reduces to $$1{+}\frac{\alpha
        }{\alpha{-}1}\,{\ydens}^{\alpha{-}1}{=}\br{1{-}\frac{\beta^2}{\alpha{-}1}}^{\!\!-1}.$$
On comparing this expression with the sonic velocity formula \eqref{eq:sonicspeed1}, it follows that $$c_s^2=\beta^2$$ on  the principal critical line, which means that this line is also a {\it sonic curve}. At a turning point the accretion velocity and the sonic speed are equal, and the density profile steepens sharply, which is characteristic of shock points. Therefore, the principal critical line can be called the {\it line of shocks} while the turning points can be called (sonic) {\it shock points}. More specifically,  represented on the $x$-$\beta$ plane, the line of shocks can be defined as $\partial_{\beta} A(u,\beta){=}0$, which gives
\begin{equation}\label{eq:shockcurve} u-\br{ 1 \!-\! \frac{\beta^2}{\alpha\!  -\! 1}       } {\sqrt{1 \!-\! \beta^2}}=0.\end{equation} 
Now, the result that $c_s{=}\beta$ on this line is seen by eliminating ${\ydens}$ from  \eqref{eq:sonicspeed1} with the help of the constraint $\Epsilon{=}\epsilon$, namely by inserting \eqref{eq:shockcurve} in $${c_s^2(u,\beta)}{\equiv}c_s^2({\ydens}(u,\beta)){=}(\alpha{-}1)\!\br{\!1{-}\frac{u}{\sqrt{1-\beta^2}}\!}\!{<}\alpha{-}1.$$
Several other  properties that can be inferred for points confined to the line of shocks are following. \eqref{eq:shockcurve} implies that $c_s{=}\beta{\to}\sqrt{\alpha{-}1}$ or $c_s{=}\beta{\to}1$ as $u{\to}0$ on this line. 
If  $\beta{\to}\sqrt{\alpha{-}1}$ (which means that $\alpha$ must not exceed $2$), then ${{\ydens}{\to}\infty}$ and the limiting $\Kappa$ can be real only for $1{<}\alpha{\leqslant}2$. Then $\Kappa{\to}0$ for $\alpha{>}\au$, $\Kappa{\to}\infty$ for $\alpha{<}\au$, or $\Kappa$ tends to a nonzero constant for $\alpha{=}\au$ (for $\alpha{=}2$, $\beta{\to}1$, $\Kappa{\to}0$, ${\ydens}{\to}\infty$). If $\beta{\to}1$, the limiting $\Kappa$ can be real only for $\alpha{>}2$, then $\Kappa{\to}0$ and  ${\ydens}{\to}(\frac{\alpha-1}{\alpha(\alpha-2)})^{1/(\alpha{-}1)}$. 
  
On the $u$-$\beta$ plane one can also distinguish a {\it conjugate critical line} $\partial_u A(u,\beta){=}0$ that makes sense only for $\alpha{>}\au$: 
\begin{equation}\label{eq:conjline} u-\frac{\alpha -\au }{\alpha-1}\sqrt{1-\beta ^2}=0,\quad \alpha{>}\au.\end{equation}
For points confined to this line, both ${\ydens}$ and $c_s$ are constant: ${\ydens}^{\alpha-1}{=}\frac{\alpha-1}{\alpha}\frac{\au-1}{\alpha-\au}$ and $c_s{=}\sqrt{\au{-}1}{<}1$. It is clear that the particular value of the index $\alpha{=}\au$ is critical for the model. Another conjugate critical line, different from the previous one, is associated with the accretion rate surface on the $u$-$c_s$ plane and reads
\begin{equation}\label{eq:conj2}\frac{c_s^2}{\alpha{-}1}{+}\frac{u}{\sqrt{2{-}\au}}=1.\end{equation}
This line is defined by the condition  $\partial_u B(u,c_s){=}0$ with a function $B(u,c_s){\equiv}K(u,\beta(u,{\ydens}(c_s)),{\ydens}(c_s))$, where ${\ydens}(c_s)$ is the inversion of the relation \eqref{eq:sonicspeed1} between ${\ydens}$ and $c_s$, and $\beta(u,{\ydens})$ is the solution of the enthalpy constraint in \eqref{eq:radial_integrals} involving variables $u,\beta,{\ydens}$. For all points confined to this line, the flow velocity is the same $\beta{=}\sqrt{\au{-}1}$. If $u{\to}0$ then $c_s{\to}\sqrt{\alpha{-}1}$ and there is a limitation $\alpha{\leqslant}2$ if the sound speed is not allowed to be superluminal.

\newlength{\psize}
\newlength{\csize}
\setlength{\psize}{1.061\columnwidth}
\setlength{\csize}{0.5\psize}
\newcommand{\ul}{\includegraphics[trim=3 12 24 3,clip,width=\psize]{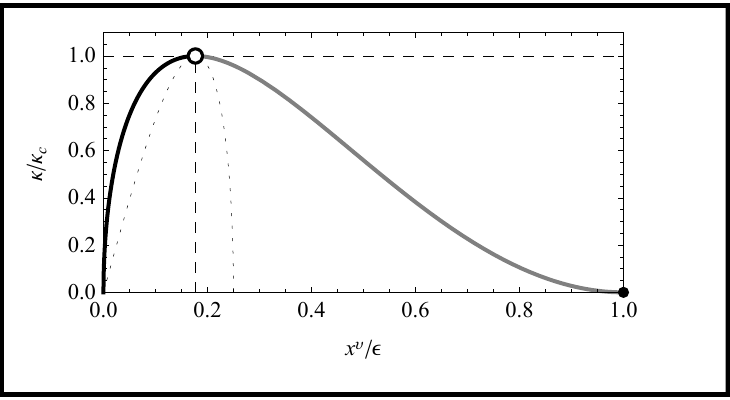}}
\newcommand{\ur}{\includegraphics[trim=3 12 24 3,clip,width=\psize]{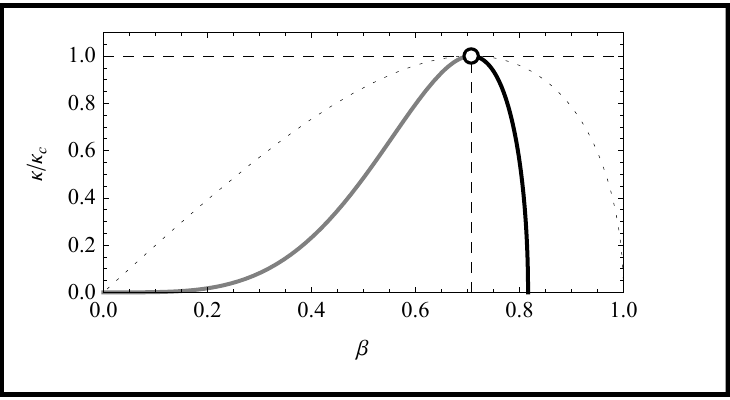}}
\newcommand{\dl}{\includegraphics[trim=3 12 24 3,clip,width=\psize]{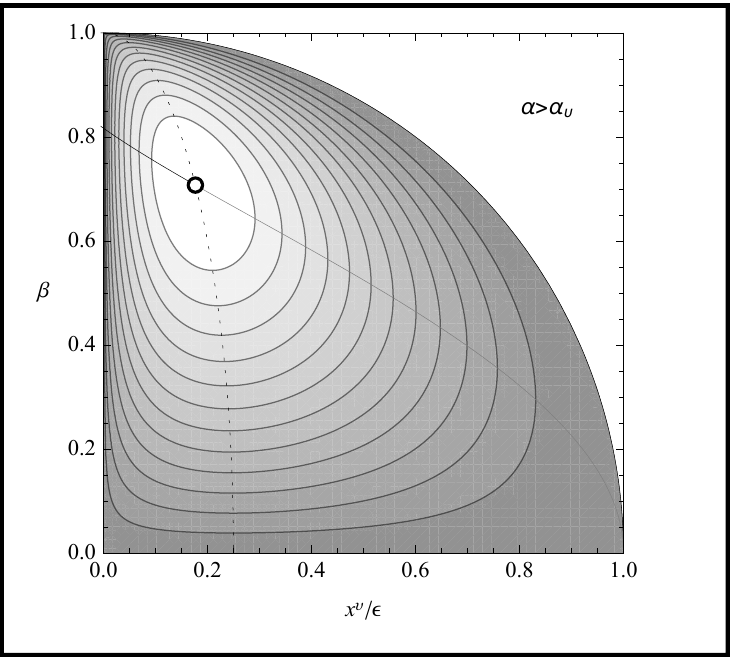}}
\newcommand{\drul}{\includegraphics[trim=3 12 24 3,clip,width=\csize]{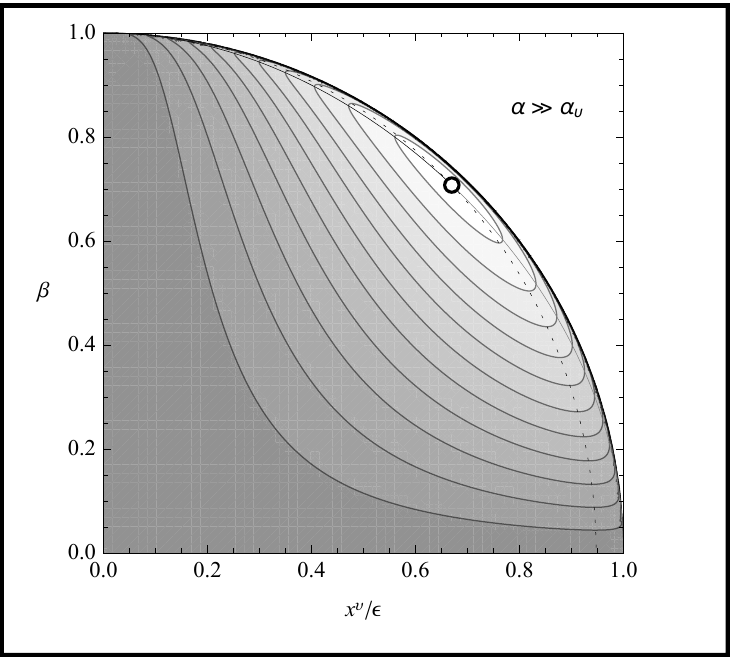}}
\newcommand{\drur}{\includegraphics[trim=3 12 24 3,clip,width=\csize]{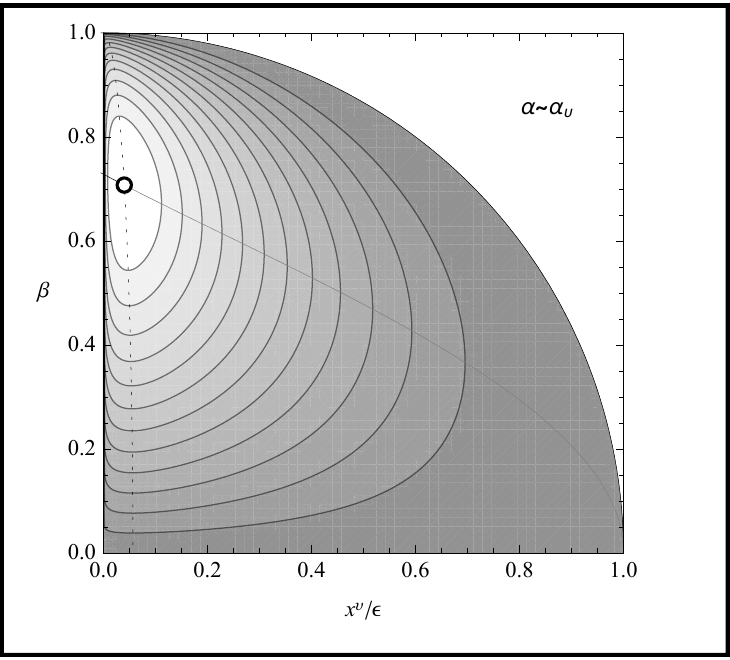}}
\newcommand{\drdl}{\includegraphics[trim=3 12 24 3,clip,width=\csize]{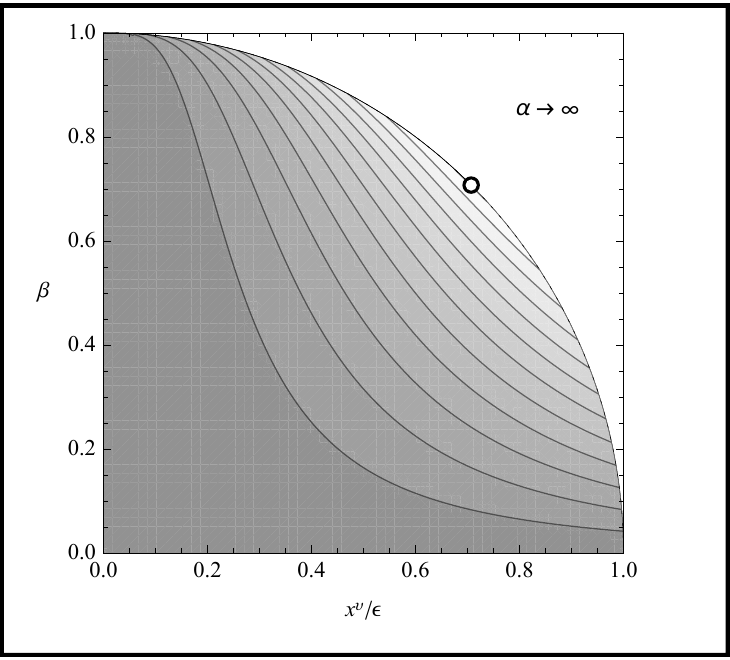}}
\newcommand{\drdr}{\includegraphics[trim=3 12 24 3,clip,width=\csize]{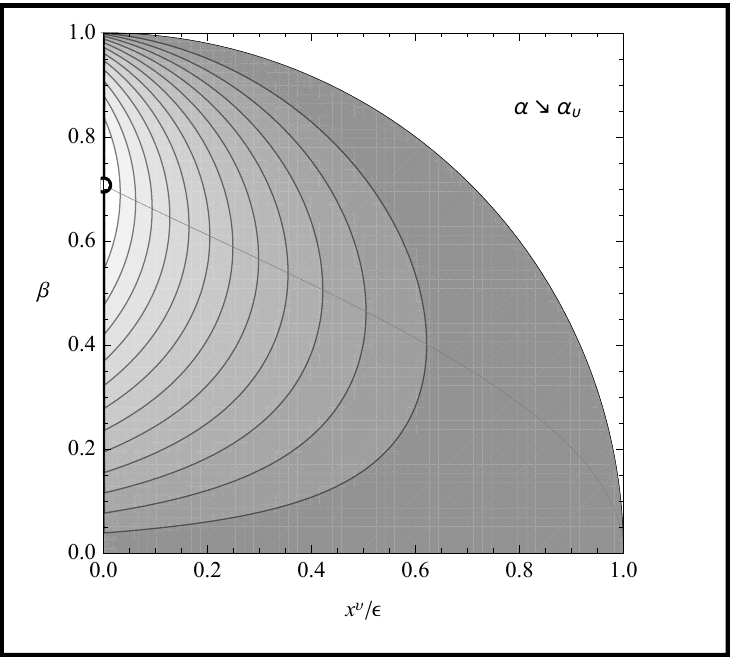}}
\newcommand{\hprba}{\includegraphics[trim=3 12 24 3,clip,width=\csize]{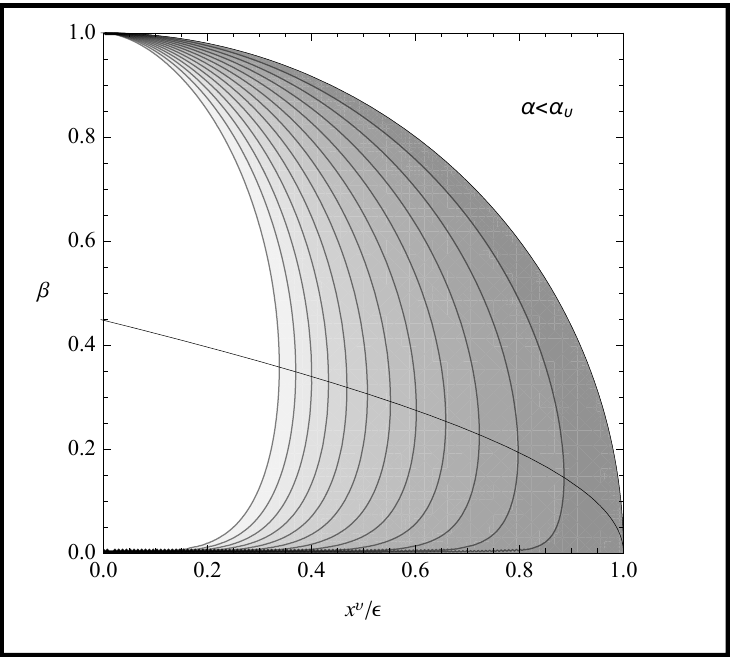}}
\newcommand{\hprbb}{\includegraphics[trim=3 12 24 3,clip,width=\csize]{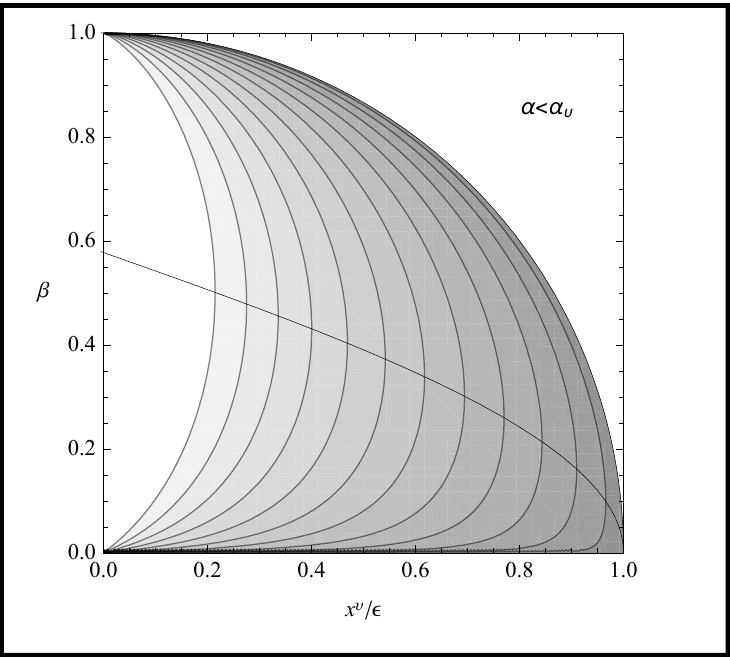}}
\newcommand{\hprbc}{\includegraphics[trim=3 12 24 3,clip,width=\csize]{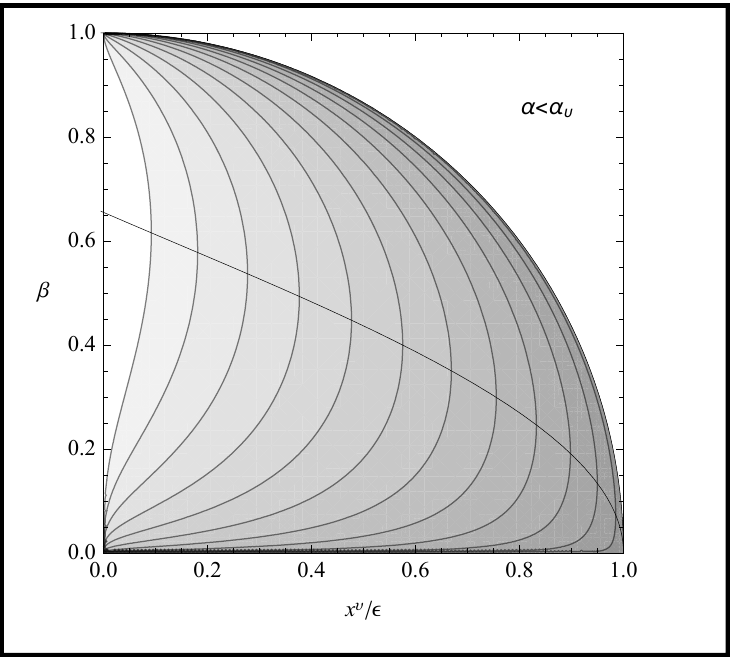}}
\newcommand{\parab}{\includegraphics[trim=3 12 24 3,clip,width=\csize]{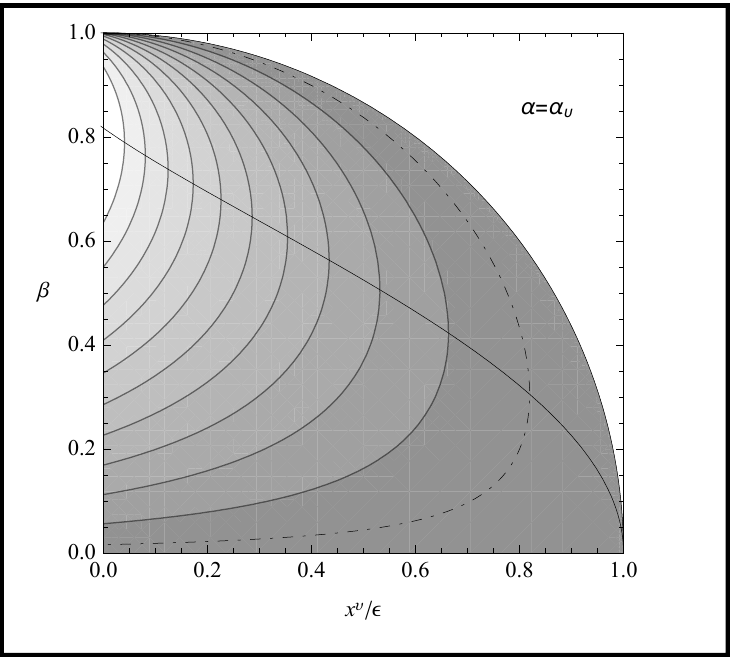}}
\begin{figure*}[t!]
\centering
\renewcommand{\arraystretch}{0.2}
\begin{tabular}{|@{}l@{}|@{}l@{}|@{}l@{}|@{}l@{}|}
\hline
\multicolumn{2}{|@{}c@{}|}{\ul} 				& \multicolumn{2}{@{}c@{}|}{\ur} \\ 
\multicolumn{2}{|@{}l@{}|}{\raisebox{0pt}{\scriptsize\phantom{i} (Ib)}} 				& \multicolumn{2}{@{}l@{}|}{\raisebox{0pt}{\scriptsize\phantom{i} (Ic)}} \\ \hline
\multicolumn{2}{|@{}c@{}|}{\raisebox{4.2cm}{\multirow{3}{*}{\dl}}} 	& \drul  &  \drur \\
\multicolumn{2}{|@{}c@{}|}{} & {\scriptsize\phantom{i} (IIb)} & {\scriptsize\phantom{i} (IIc)} \\
\cline{3-4} \multicolumn{2}{|@{}c@{}|}{}				& \drdl  &  \drdr \\ 
\multicolumn{2}{|@{}l@{}|}{{\scriptsize\phantom{i} (Ia)}}	& {\scriptsize\phantom{i} (IIa)} & {\scriptsize\phantom{i} (IId)}\\ \hline
 \hprba & \hprbb      					& \hprbc  &  \parab \\ 
 {\scriptsize\phantom{i} (IVc)}&{\scriptsize\phantom{i} (IVb)}&{\scriptsize\phantom{i} (IVa)}&{\scriptsize\phantom{i} (III)}\\\hline							
\end{tabular}
\caption{ Radial accretion solutions shown as  level lines of the accretion rate surface.  Ia) A typical diagram with the elliptic point ($\alpha{>}\au$) (shown for the example values $\alpha{=}5/3$ and $\au{=}3/2$) compared with the diagrams of critical and limiting values of $\alpha$: (IIa) $\alpha{\to}\infty$, (IIb) $\alpha{\gg}\au$, (IIc) $\alpha{\sim}\au$ and (IId) $\alpha{\searrow}\au$. The line of shocks \eqref{eq:shockcurve} (black and gray thin lines) crosses each level line at two respective shock points and separates it to the supersonic and the subsonic branch. The conjugate critical line \eqref{eq:conjline} (dotted line) intersects each supersonic branch at the maximum velocity position and each subsonic branch at the minimum velocity position. The stationary point is shown with small black circle in each panel. Corresponding to diagram Ia, the cross-sections of the accretion rate surface along the line of shocks and along the conjugate critical line  are shown as a function of position (Ib) or as a function of velocity (Ic). Solutions for the critical value $\alpha{=}\au$ are shown in panel III (here $\au{=}5/3$);  an example analytical solution is shown with the dot-dashed line. For $\alpha{<}\au$ one can distinguish three types of level lines with various analytic properties at $x=0$ (here $\au{=}3/2$).}
\label{fig:1}
\end{figure*}

\setlength{\psize}{0.68\columnwidth}
\newcommand{\Ghb}{\includegraphics[trim=3 12 24 3,clip,width=\psize]{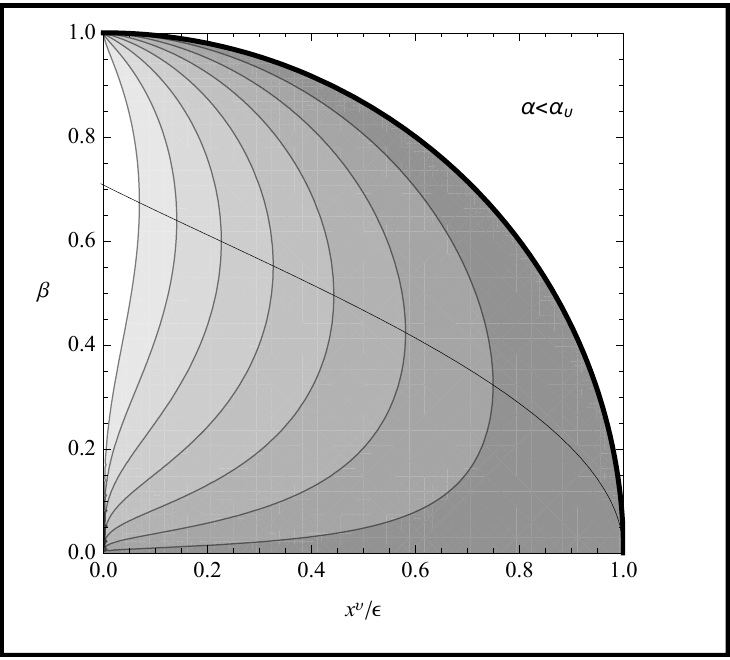}}
\newcommand{\Ghn}{\includegraphics[trim=3 12 24 3,clip,width=\psize]{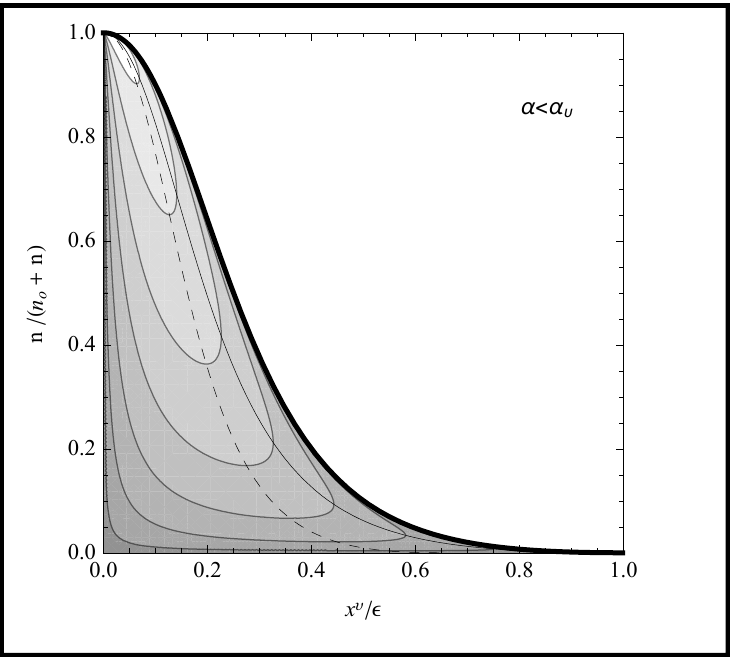}}
\newcommand{\Ght}{\includegraphics[trim=3 12 24 3,clip,width=\psize]{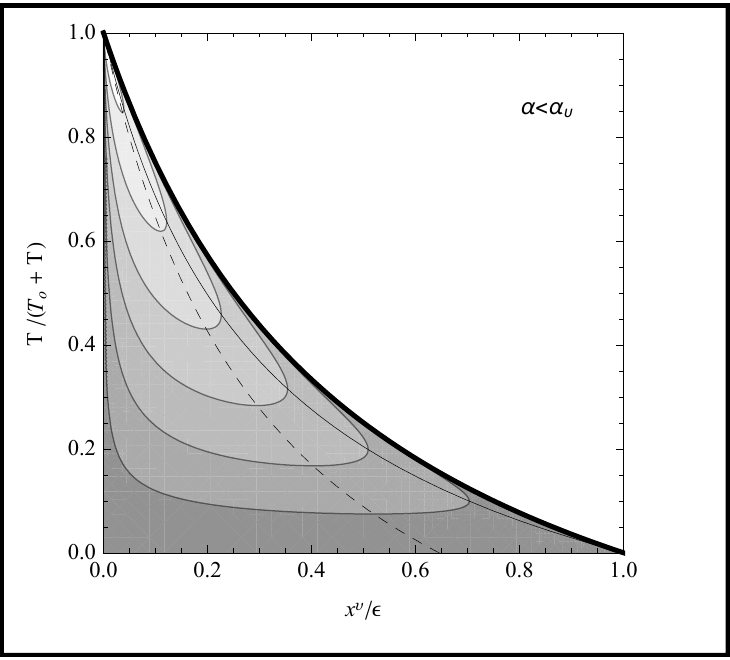}}
\newcommand{\Gpb}{\includegraphics[trim=3 12 24 3,clip,width=\psize]{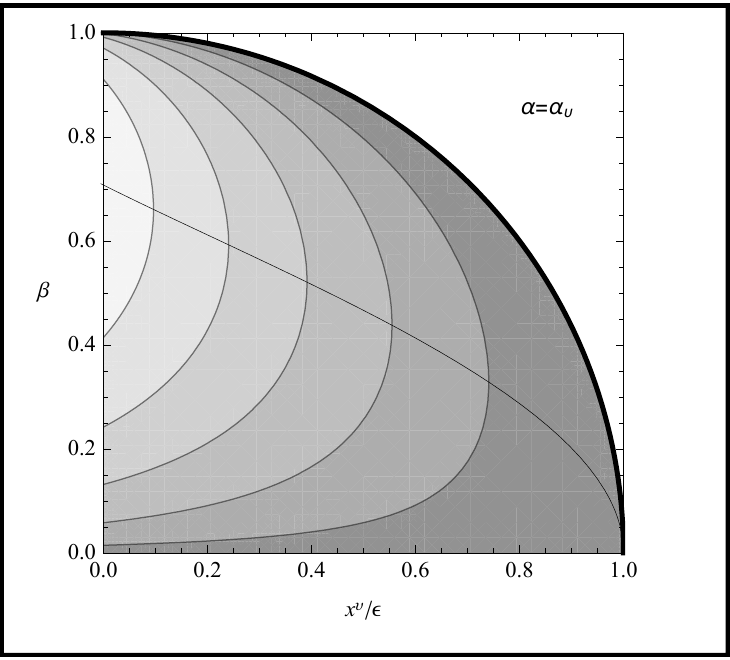}}
\newcommand{\Gpn}{\includegraphics[trim=3 12 24 3,clip,width=\psize]{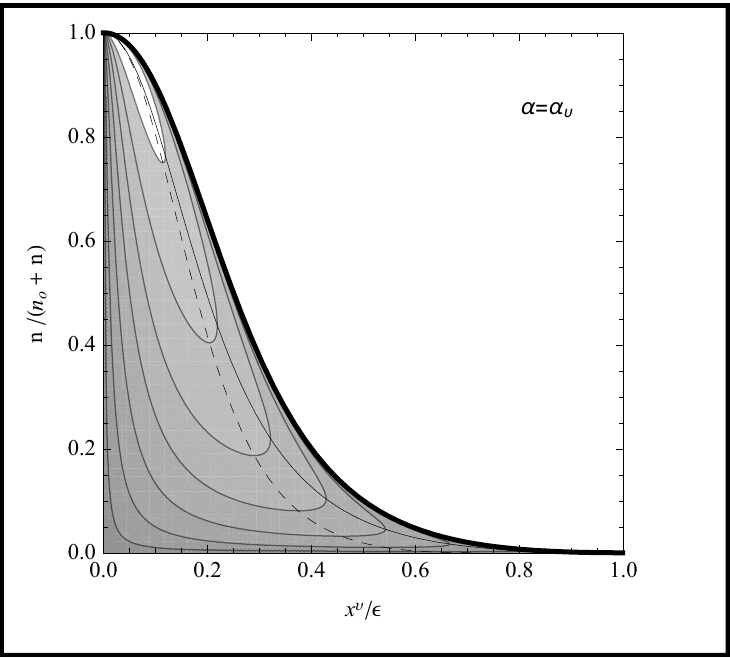}}
\newcommand{\Gpt}{\includegraphics[trim=3 12 24 3,clip,width=\psize]{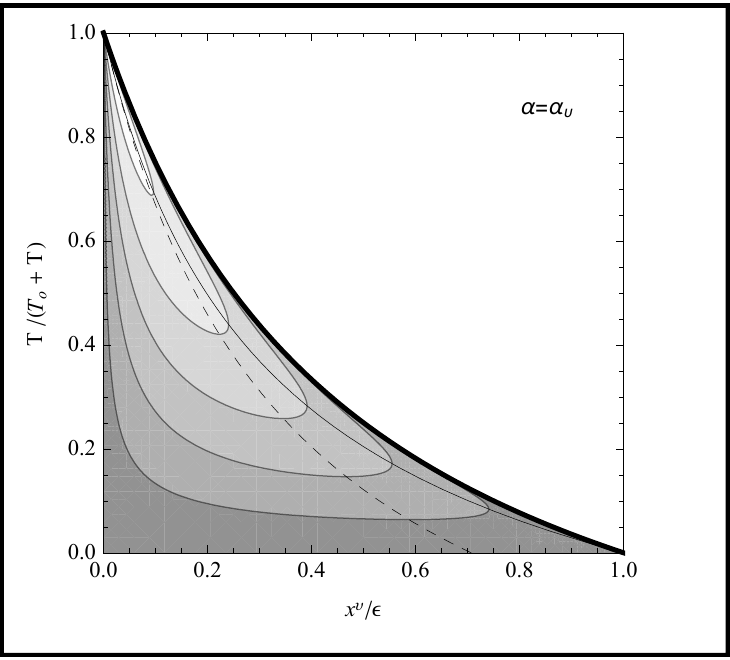}}
\newcommand{\Geb}{\includegraphics[trim=3 12 24 3,clip,width=\psize]{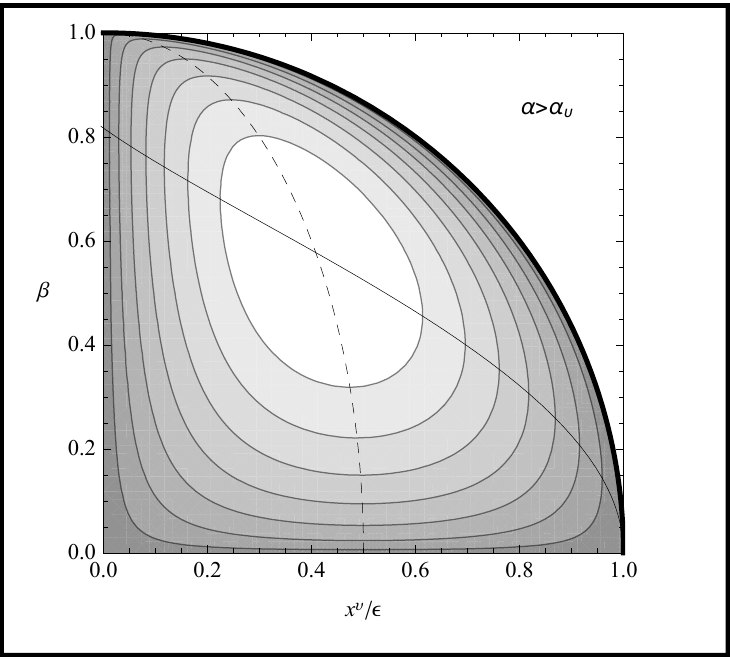}}
\newcommand{\Gen}{\includegraphics[trim=3 12 24 3,clip,width=\psize]{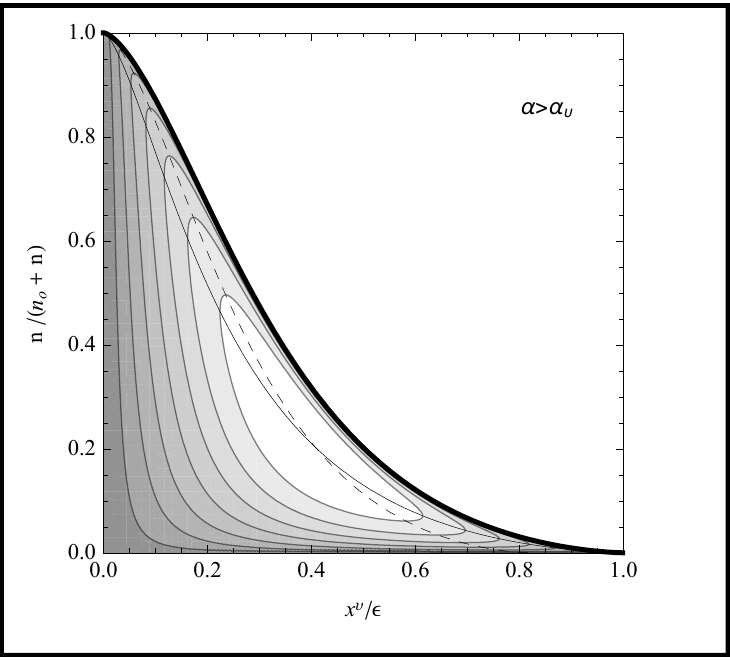}}
\newcommand{\Get}{\includegraphics[trim=3 12 24 3,clip,width=\psize]{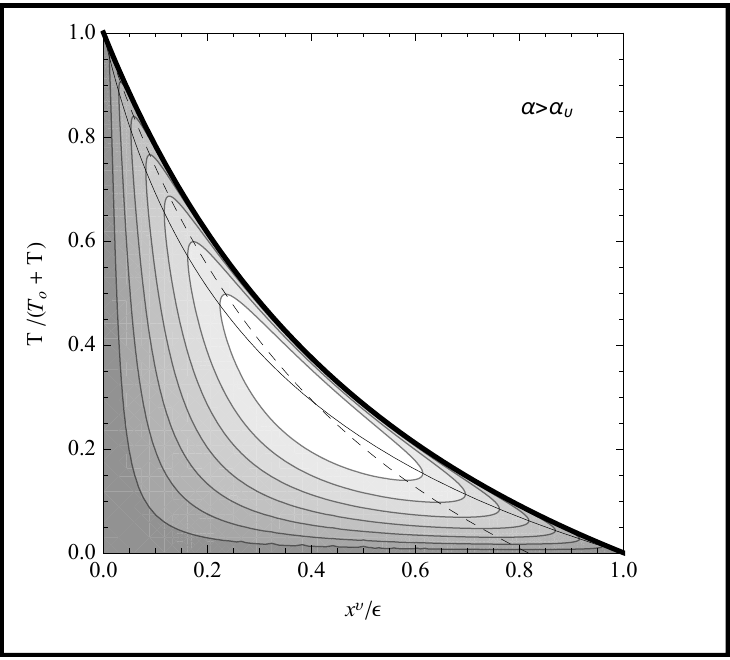}}
\begin{figure*}[t!]
\centering
\renewcommand{\arraystretch}{0.2}
\begin{tabular}{|l@{}|l@{}|l@{}|}
\hline
\Geb & \Gpb & \Ghb \\ 
 {\scriptsize\phantom{i} 1a} & {\scriptsize\phantom{i} 2a} & {\scriptsize\phantom{i} 3a}\\ \hline
\Gen & \Gpn & \Ghn \\ 
 {\scriptsize\phantom{i} 1b} & {\scriptsize\phantom{i} 2b} & {\scriptsize\phantom{i} 3b}\\ \hline
\Get & \Gpt & \Ght \\ 
 {\scriptsize\phantom{i} 1c} & {\scriptsize\phantom{i} 2c} & {\scriptsize\phantom{i} 3c}\\
\hline
\end{tabular}
\caption{Typical diagrams for radial accretion solutions onto Levi-Civita string, shown as level lines of the accretion rate surface on the position-velocity plane  ({\it row a}), on the position-number density plane  ({\it row b}), and on the position-temperature plane ({\it row c}), for three qualitatively distinct regimes:  $\alpha{>}\au$ ({\it column 1}), $\alpha{=}\au$ ({\it column 2}), and $\alpha{<}\au$  ({\it column 3}) /the example values shown are following $3/4{<}\alpha{<}\au{<}5/3$, $3/2{=}\alpha{=}\au{<}5/3$, and $5/3{=}\alpha{>}\au{=}4/3$.
     The line of shocks  (crossing all level lines at shock points where the slope of a level line diverges as a function of position) is plotted with black thin line.  The conjugate critical line (intersecting the level lines at their minima or maxima) is plotted with dotted line. }
\label{fig:2}
\end{figure*}

\subsection{Three accretion regimes}

In this section, the upper bound $\alpha{<}2$, which agrees with the Taub limit referred to in \secref{sec:taub}, is assumed. The limit also follows from the discussion of the previous
section, where the number $\sqrt{\alpha{-}1}$ along with $\sqrt{\au{-}1}$ appeared as critical speeds (in units of the speed of light) both for the matter flow and for the propagation of sound waves.

Some properties of the accretion solutions regarded as level lines can be established by investigating the geometry of the accretion rate surface \eqref{eq:accrsurface}.

\medskip
\noindent
$\bullet$ For $\alpha{>}\au $, the surface has a single stationary point at the intersection of the line of shocks \eqref{eq:shockcurve} with the conjugate critical line \eqref{eq:conjline}. At this point 
$$\uc{=}\frac{\alpha{-}\au}{\alpha{-1}}\sqrt{2{-}\au},\,\,
\theta_{\!c}{=}\yc^{\alpha-1}{=}
\frac{\alpha{-}1}{\alpha}{\cdot}
\frac{\au{-}1}{\alpha{-}\au},\,\,
\bc{=}\sqrt{\au{-}1}$$
with $c_s{=}\bc$. Evaluated at that point, both eigenvalues of the Hessian 
$\partial^2_{ij}A$ are negative. This observation follows from examining the signs of the main minors of the Hessian matrix at the stationary point:  
        \begin{eqnarray*}\left.\frac{u^2 \beta^2\,\mathrm{det}\!\br{\partial^2_{ij}A(u,\beta)}}{2\,A^2(u,\beta)}\right|_c&=&
  \frac{\alpha {-}\au  }{(2{-}\au ) (\au  {-}1)^2}\gt0\\
    \left.\frac{u^2\,\partial^2_uA(u,\beta)}{A(u,\beta)}\right|_c&=& -\frac{ \alpha{ -}\au }{\left(\au{-}1\right){}^2}\lt0.\end{eqnarray*}
        The stationary point is thus an elliptic point and a local maximum. 
    
    Accordingly, the level lines are closed loops encircling the stationary point. Each level line consists of a subsonic and a supersonic branch with their endpoints joined at the intersection of the loops with the line of shocks.  In the limit $\varkappa{\to}0$ the supersonic branch overlaps with the boundary line $u{=}\sqrt{1{-}\beta^2}$ of free fall from rest at $u{=}1$.
As $\varkappa$ increases, the loops shrink away to the stationary point and there are no solutions for $\varkappa{>}A(\uc,\bc)$.  

Parameterized locally as $u(\beta)$, the level lines are concave down or concave up at shock points, respectively, to the right or to the left of the stationary point $\uc$
(the shock points are turning points for the solutions regarded as functions $\beta(u)$). This follows from the fact that $u'{=}0$  and hence $u''{=}{-}\partial^2_{\beta}A/\partial_{u}A$ at shock points, the expression being positive if $\beta{>}\bc$ ($u{<}\uc$) and negative if $\beta{<}\bc$ ($u{>}\uc$). An example solution (for $\au{=}3/2$ or  $\upsilon{=}(\sqrt{5}{-}1)/2$ -- the golden ratio conjugate -- and for $\alpha{=}5/3$) is shown in panel Ia in \figref{fig:1}. The position-velocity coordinates of shock points and the extremum velocity points as functions of the accretion rate can be read off from the parametric plots Ib and Ic.   
In the formal limit $\alpha{\gg}1$ (see panels IIa, IIb) the line of shocks and the conjugate critical line both converge to the boundary line  $u{=}\sqrt{1{-}\beta^2}$  (cf., \eqref{eq:shockcurve} and \eqref{eq:conjline}), with $\uc$ increasing toward $\sqrt{2{-}\au}{<}1$, while the solutions converge in this limit to $\beta(u){=}(1{+}(u^{\frac{1}{\au{-}1}}{/}\varkappa)^2)^{-1/2}$ (inferred from the equation $A(u,\beta){=}\varkappa$ in this limit). In the opposite limit $\alpha{\searrow}\au$  (see, panels IIc and IId), $\uc{\to}0$ and the conjugate critical line overlaps with the $u{=}0$ line. The limiting solution will be discussed in some more detail below. 

\medskip

\noindent $\bullet$ For $\alpha{=}\au$ the solutions can be found in exact form $u(\beta){=}\sqrt{1{-}\beta^2}{-}(\varkappa\,\beta^{-1}\sqrt{1{-}\beta^2})^{\au{-}1}$ defined for $\beta$ in a $\varkappa$-dependent region where $u(\beta){>}0$.
The accretion rate attains its maximum at $u{=}0$, $\beta{=}\sqrt{\au{-}1}$, however  there is no a stationary point. The slope of a level line at $u{=}0$ is finite $u'(\beta){=}\frac{\au-1-\beta^2}{\beta\sqrt{1-\beta^2}}|_{u(\beta)=0}$ -- it is positive for the subsonic branch of the line (for which $0{<}\beta{<}\sqrt{\au{-}1}$ at $u{=}0$) and negative for the supersonic branch (for which  $\sqrt{\au{-}1}{<}\beta{<}1$ at $u{=}0$). 
The two branches meet at the shock point lying on the line of shocks \eqref{eq:shockcurve}. The level lines parameterized locally as $u(\beta)$ are concave down at the shock points at which $u'{=}0$, hence $u''{=}{-}\partial^2_{\beta}A/\partial_{u}A$ at these points, and evaluated on the line of shocks the latter expression can be shown to be negative. Thus, the shock points are right turning points for the considered solutions. In the limit of vanishing accretion rate the solutions tend to the boundary line $u{=}\sqrt{1{-}\beta^2}{\leqslant}1$ of free fall. An  example phase diagram together with an exact solution indicated with the dot-dashed line are shown in panel III of \figref{fig:1}.
 
\medskip

\noindent
$\bullet$ For $\alpha{<}\au $ there is no a stationary point. The accretion rate can be arbitrarily high.
 As previously, each level line consists of a subsonic branch and a supersonic branch, however with different asimptotics. In the limit $u{\searrow}0$ the asymptotics is $\beta{\sim}a\,u^s$ with $s{=}\frac{\au-\alpha}{(\alpha-1)(\au-1)}$ and $a{=}{\frac{\kappa}{ \sqrt[\au{-}1]{\epsilon}  }}\! \sqrt[\alpha{-}1]{\!\frac{\alpha}{\alpha{-}1}\!}$ for the subsonic branch, 
 and $\beta{\sim}1{-}b\,u^t$ with $0{<}t{=}2\frac{\au{-}\alpha}{(2{-}\alpha)(\au{-}1)}{<}2$, $b{=}\frac{1}{2}a^{\frac{2(\alpha{-}1)}{2{-}\alpha}}$
for the supersonic branch. The two branches converge at the sonic shock point located on the line of shocks \eqref{eq:shockcurve}. For $\alpha{<}\au$ the level lines parameterized locally as $u(\beta)$ are always concave down at the shock points, and this can be shown the same way as previously. Evaluated on the line of shocks, $u''$ can be shown to be negative if $\alpha{<}2$ and $\alpha{<}\au$. Thus, the shock points are right turning points for the considered solutions. The solutions become identical to the boundary line $u{=}\sqrt{1{-}\beta^2}{<}1$ of free fall in the limit $\varkappa{\to}0$, while in the opposite limit $\varkappa{\to}\infty$ the solutions overlap with the line $u{=}0$. The  example three types of level lines with various analytic properties at $x{=}0$ are shown in panel IV of \figref{fig:1}. The behavior of other thermodynamical quantities is shown in figure \figref{fig:2}.

\section{Physical discussion}

To discus the physical properties of the accretion,  it is convenient to refer physical quantities to an initial state specified at some arbitrary radius $r{=}\rfty$ by the inward radial velocity $\bfty$ (in units of the speed of light),  the proper number density $n_{\infty}$, and the parameter of the initial temperature $T_{\infty}$  $$\tfty{\equiv}\frac{k_B\, T_{\infty}}{m\,c^2}.$$  Then the two independent functions of state $\beta$ and $${\ydens}\equiv n/n_{\infty},$$ at any other radius $r$ (or any two combinations of them) will be functions of the initial state parameters and of the ratio $$x\equiv r/\rfty,$$ while  $x{=}1$ and ${\ydens}{=}1$ will correspond to the initial state. The remaining thermodynamical quantities such as the temperature, pressure or the speed of sound (in units of the speed of light) can be expressed as functions of the density (cf. \secref{sec:taub}) $$\theta{=}\theta_{\infty}{\ydens}^{\alpha{-}1},\quad
\frac{p}{{n_{\infty}mc^2}}{=}\theta_{\infty}{\ydens}^{\alpha},$$
$$\frac{e}{n_{\infty}mc^2}{=}{\ydens}+\chi\theta_{\infty}{\ydens}^{\alpha},\quad
c_s^2{=}\frac{
\alpha \theta_{\infty}{\ydens}^{\alpha{-}1}
}{
1{+}\frac{\alpha}{\alpha{-}1}\theta_{\infty}{\ydens}^{\alpha{-}1}.
}$$
In this notation the corresponding critical values at the stationary point are
$$
\frac{\xc^{\upsilon}}{\epsilon_{\infty}}{=}\frac{\alpha{-}\au}{\alpha{-1}}\sqrt{2{-}\au}{<}1,\qquad
\theta_c{=}\theta_{\infty}\yc^{\alpha-1}{=}\frac{\alpha{-}1}{\alpha}{\cdot}
\frac{\au{-}1}{\alpha{-}\au},$$
$$\bc{=}\sqrt{\au{-}1},\qquad
\kc{=}\sqrt{\frac{\au{-}1}{2{-}\au}}\! \left(\!\frac{\theta_c}{\theta_{\infty}}\!\right)^{\!\frac{1}{\alpha{-}1}} \!\!\!\left(\xc\right)^{\!\frac{\upsilon}{\au{-}1}}.
$$
As already described at the beginning of \secref{sec:qualitres}, the structure of the original integrals of motion \eqref{eq:1stinteg1and2} and \eqref{eq:1stinteg3} allowed to cancel out the length parameter $\alen$ introduced by the metric \eqref{eq:metricab} and to use arbitrary reference density. This cancellation can be done by equating values of each integral calculated for physical states from the same solution curve at two distinct radii $r$ and $\rfty$. The integrals can be then recast to forms 
involving only parameters $Q$, $\alpha$, $\tfty$, functions of state $\psi$ and $\beta$ at these radii and the ratio $r/\rfty$. Unlike for $\alen$, the role of the parameter $Q$ is essential in the accretion context, since functions of state depend on $Q$ in a nontrivial way. The parameter $Q$ can be measured by means of frequency shifts, as already described in \secref{sec:spacetime}. However, since the form of the integrals is independent upon the assumption $\alen{=}\blen$, one can assume $Q{=}\upsilon$ as if one had started with the Wilson metric \eqref{eq:metric} from the beginning. Hence, the integrals  corresponding to  \eqref{eq:1stinteg1and2} and \eqref{eq:1stinteg3}  read\footnote{The equations \eqref{eq:epsfty} and \eqref{eq:kapfty} are analogous to \eqref{eq:1stinteg1and2} and \eqref{eq:1stinteg3}. The following difference in the notational conventions should be
stressed:  in \eqref{eq:epsfty} and \eqref{eq:kapfty}, $x{=}1$ corresponded to the length parameter $\alen$ in the line element \eqref{eq:metricab}, while ${\ydens}{=}1$ corresponded to the density $n_o$ at a reference temperature $T_o{=}mc^2/k_B$. Also the units of specific enthalpy and accretion constants are different here, however their actual values are not needed. }
\begin{equation}\label{eq:epsfty} 
\frac{1{+}\frac{\alpha}{\alpha{-}1}\theta_{\infty}{\ydens}^{\alpha{-}1}}{\sqrt{1{-}\beta^2}}x^{\upsilon}{=}\frac{1{+}\frac{\alpha}{\alpha{-}1}\theta_{\infty}}{\sqrt{1{-}\beta_{\infty}^2}}{\equiv}\epsilon_{\infty}{>}1
\end{equation}
\begin{equation}\label{eq:kapfty}
\frac{\beta\,{\ydens}\,x^{\frac{\upsilon}{\au{-}1}}}{\sqrt{1{-}\beta^2}}{=}\frac{\beta_{\infty}}{\sqrt{1{-}\beta_{\infty}^2}}{\equiv}\kappa_{\infty}{>}\beta_{\infty}.\end{equation}
The length parameters in the metric \eqref{eq:metricab} would play a role if one were interested in the questions involving physical distances or volumes. For example,  what is a geodesic distance of a particular state at a given circumferential radius $r$, or, given the predicted $n(r)$, what is the actual number of particles between cylinders separated by some $\Delta r$. The problem of distance is solved once the interior metric is known or the ratio $\alen/\blen$ has been determined from independent measurements. The second problem can be  resolved without knowing the actual value of the ratio by making use of the symmetry of the equations with respect to rescaling the number density by a constant factor.  In place of the true value $n_{\infty}$ one can introduce the apparent number density $n_{\infty}'$ defined at $r{=}\rfty$ based on the Euclidean formula $n'_{\infty}{=}N_{\infty}/(2\pi \rfty\Delta r\Delta z)$. Here,  $N_{\infty}$ is the true number of particles obtained by counting the particles in the physical volume enclosed by coordinate segments  $\Delta r$ and $\Delta z$ (to first order in $\Delta r$ the volume at $\rfty$ for metric \eqref{eq:metricab}  is 
 $2\pi \rfty (\blen/\alen)^{\frac{Q^2}{1{+}Q}} (\blen/\rfty)^{\frac{Q(1{-}Q)}{1{+}Q}}\Delta r\Delta z$). 

\subsection{\label{sec:chpts}Characteristic points of the accretion solutions}

The parametrization of the line of shocks on the $x$-$\beta$ plane is obtained by solving the energy integral for $x$ and eliminating $\theta_{\infty}{\ydens}^{\alpha{-}1}$ with the help of  \eqref{eq:shock_line_def}:
\begin{equation}\label{eq:shock_line2}\frac{(x(\beta))^{\upsilon}}{\epsilon_{\infty}}{=}\br{1{-}\frac{\beta^2}{\alpha{-}1}}\sqrt{1{-}\beta^2}\end{equation} (the result is analogous to \eqref{eq:shockcurve}), while the accretion rate and density regarded as functions of $\beta$ on this line read 
$$\kappa(\beta){=}\frac{\beta {\ydens}(\beta) (x(\beta))^{\frac{\upsilon}{\au{-1}}}}{\sqrt{1{-}\beta^2}}, \quad 
\theta_{\infty}{\cdot}({\ydens}(\beta))^{\alpha{-}1}{=}\frac{1}{\alpha}\frac{\beta^2}{1{-}\frac{\beta^2}{\alpha{-}1}}.$$
Now, to get some insight into the behavior of physical quantities, one can study position of some characteristic points on the level lines of accretion solutions as functions of the initial data $\kapfty$, $\epsfty$ and $\tfty$. This will be done by assuming $$1{<}\au{<}\alpha{<}2,\qquad 0{<}\upsilon{<}1,$$ in accordance with what has been established in previous sections. The example level lines and the characteristic points are illustrated in \figref{fig:phys2} consisting of two diagrams that completely describe the independent mechanical and thermodynamical functions of state in terms of the flow velocity and the speed of sound. 

\newlength{\figsize}
\setlength{\figsize}{0.88\columnwidth}
\begin{figure*}[t!]
\centering
\begin{tabular}{@{}c@{}c@{}}
\includegraphics[trim=0 0 0 0,clip,width=\figsize]{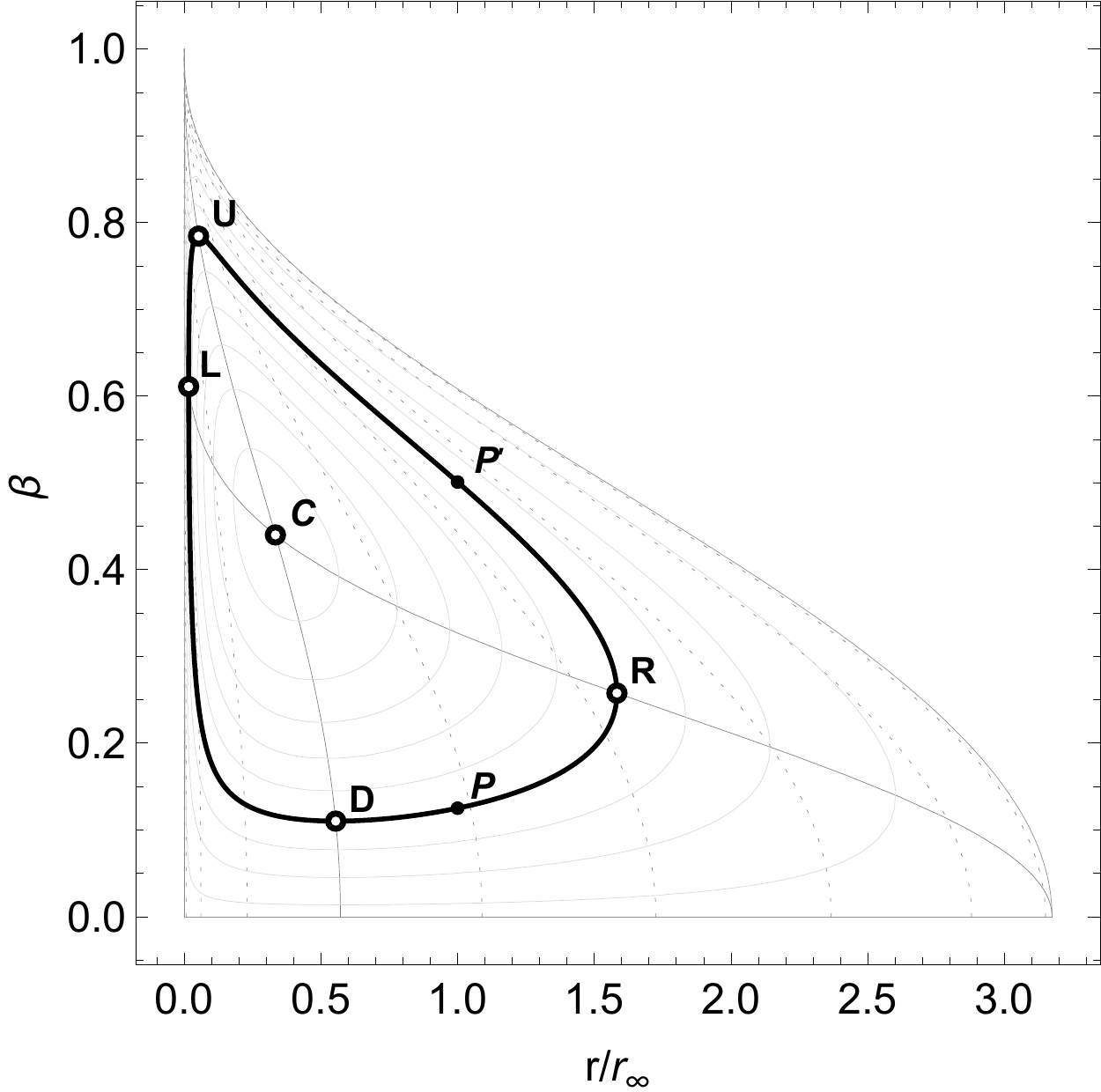}&
\includegraphics[trim=0 0 0 0,clip,width=\figsize]{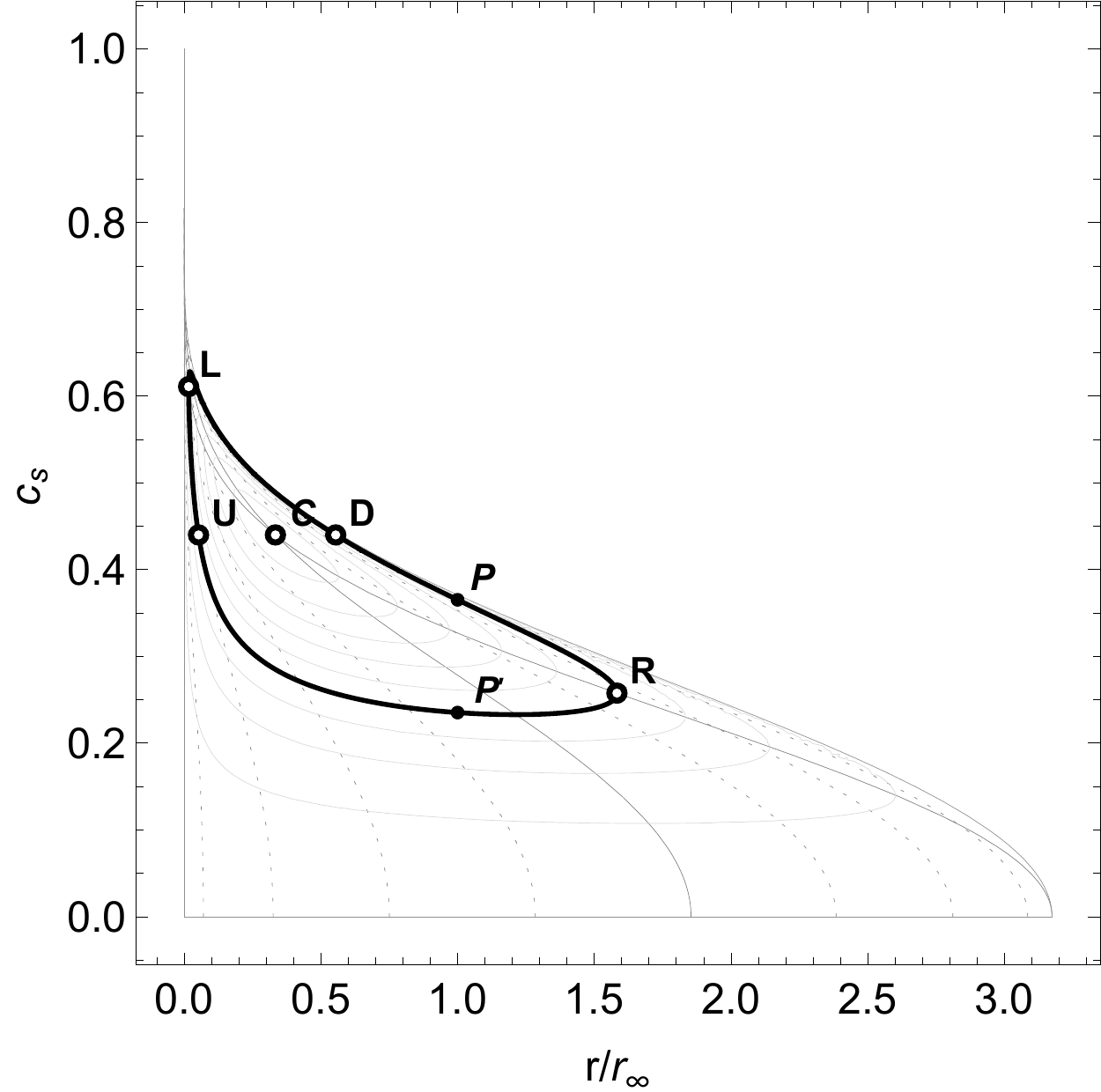}\\
\end{tabular}
\caption{\label{fig:phys2} Characteristic points on contour lines of example accretion rate surface for $1{<}\au{<}\alpha{<}2$, shown in {\it position-flow velocity} plane (top panel) and in {\it position-sonic speed} plane (bottom panel).  The central point C is the stationary point with critical (here maximum) accretion rate. The thin closed loops represent the contour lines of constant accretion rate.
The thick solid closed loops represent particular solutions through some example initial state {{(\it black dot} P)} on the subsonic branch R-D-L. The state is defined at a radius $r_{\infty}$ by temperature $T_{\infty}$, number density $n_{\infty}$ and flow velocity $\beta_{\infty}$. The line also passes through the corresponding initial state on the supersonic branch R-U-L {\it(black dot $\text{P}'$)} defined at the same radius $\rfty$ by some other parameters $T_{\infty}'$, $n'_{\infty}{=}(T_{\infty}'/T_{\infty}\!)^{\frac{1}{\alpha{-}1}}{\cdot}n_{\infty}$ and $\beta_{\infty}'$ (determined by the initial state P). 
In both panels, the shock points L  and R are joined with the principal critical line  L-C-R (or sonic line of shocks \eqref{eq:shock_line2}) delineating regions of supersonic and subsonic accretion. Points U and D represent states with the highest and the lowest flow-velocity.  In the top panel,  the U-C-D solid line is the conjugate critical line \eqref{eq:conjline2} (on which the sonic speed is constant and equal $\sqrt{\au{-}1}$). In the bottom panel, the other thin line through C, joining points of maximum and minimum sonic speed on the solution contour line, is the locus of points of the same flow velocity equal $\sqrt{\au{-}1}$ (conjugate critical line \eqref{eq:conj2}). The dotted curves are, respectively, in the upper panel -- the  lines of constant sonic speed (or constant density, temperature or pressure), and in the lower panel -- the lines of constant flow velocity.}
\end{figure*}

In the limit $\kappa_{\infty}{\to}0$, the supersonic branch of level lines tends  to the line $x^{\upsilon}{=}\epsfty\sqrt{1{-}\beta^2}$  which describes free fall from rest at $r{=}\rfty$, while the subsonic branch tends in this limit to the $\beta{=}0$ line. In this limit there are three roots of equation $\kappa(\beta){=}\kappa_{\infty}$, however, for $\alpha{<}2$ only two yield non-negative ${\ydens}$, namely, $\beta{=}0$ and $\beta{=}\sqrt{\alpha{-}1}$. For higher $\kappa_{\infty}$, the roots are close to these limiting  values  and can be represented as generalized power series in $\kappa_{\infty}$. The two roots correspond, respectively, to the right and to the left (turning) shock point, called here R and L, connecting the subsonic and the supersonic branch of the level line). 
Furthermore, each branch intersects with the conjugate critical line 
\begin{equation}\label{eq:conjline2}\frac{(x(\beta))^{\upsilon}}{\epsilon_{\infty}}{=}\frac{\alpha{-}\au}{\alpha{-}1}\sqrt{1{-}\beta^2}\end{equation} (see, \eqref{eq:conjline}). On this line $\kappa$ changes with $\beta$ according to the law $$\frac{\kappa^2(\beta)}{\kc^2}{=}
\frac{\beta^2}{\bc^2}
\br{
{\frac{1{-}\beta ^2}{1{-}\bc^2}}
   }^{\frac{1}{\upsilon (1{+}\upsilon)}}, \quad 
{\ydens}(\beta){=}\yc,$$ while ${\ydens}$  remains constant (then $c_s{=}\bc$ and $\theta{=}\theta_c$).

The intersection points are two additional characteristic points:  one called D (with the lowest flow velocity on the subsonic branch) and the other called U (with the highest flow velocity on the supersonic branch), both velocities being roots of the equation $\kappa(\beta){=}\kappa_{\infty}$. In the limit $\kappa_{\infty}{\to}0$ the roots are $0$ and $1$, respectively, and similarly as before, for higher $\kappa_{\infty}$ they can be represented as generalized power series in $\kappa_{\infty}$. 

\newcommand{\mytau}{\tau_{\text{p}}}   
\begin{table*}[]
\centering
\begin{tabular}{|c|ccc|c|}
\hline
$\text{p}$ & $x_{\text{p}}$ & $\beta_{\text{p}}$ & ${\ydens}_{\text{p}}$ & $\mytau$\\
\hline
{L} &
$ \br{\epsilon_{\infty}\sqrt{2{-}\alpha}{\cdot}\mytau}^{1/\upsilon} $ &
$ \sqrt{\alpha{-}1} \br{1{-}\frac{\mytau}{2}} $ &
$ \left(\!\frac{\alpha {-}1}{\alpha \theta_{ \infty}  {\cdot}\mytau}\right)^{\!\frac{1}{\alpha -1}} $ &
$\frac{\sqrt{\alpha{-}1}^{-\frac{(\alpha{+}1) (\au{-}1)}{\alpha{-}\au}} 
  }{\sqrt{2{-}\alpha }^{\frac{(\alpha{-}1) (2{-}\au )}{\alpha{-}\au }}} 
   \sq{\alpha\theta_{\infty}\br{\frac{\kappa_{\infty}}{{\epsilon_{\infty}}^{\frac{1}{\au{-}1}}}}^{\alpha{-1}}}^{\frac{\au -1}{\alpha{-}\au}}$\\  
{D} &
$\frac{\xc}{\sqrt[2\upsilon]{2{-}\au}} \left(1{-}\frac{\bc^2 \mytau^2}{2}\right)^{1/\upsilon }$ &
$\sqrt{\au{-}1}\cdot\mytau$ &
$\yc$ &
$\left(\sqrt[2\upsilon]{2{-}\au}\right)^{\frac{1}{1+\upsilon }} \frac{\kappa_{\infty}}{\kc}$ \\
{U} & 
$x_c{\cdot}{\mytau}$  &
$1{-}\frac{2{-}\au}{2}{\cdot}{\mytau}^{2\upsilon}$ &
$\yc$ &
$\br{\frac{\kappa_{\infty}}{\kappa_c}\sqrt{\au-1}}^{1+\upsilon}$ \\
{R} &
$ \left(1{-}\frac{\alpha {+}1 }{\alpha {-}1}{\cdot}\frac{{\mytau}^2}{2}\right)^{\!\!1/\upsilon } \!\!{\epsilon_{\infty}} ^{1/\upsilon} $ &
$\mytau$ &
$ \left(\!\frac{{\mytau}^2}{\alpha\,  \theta_{ \infty} }\!\right)^{\!\!\frac{1}{\alpha -1}} $ &
$ \sq{\alpha\theta_{\infty}\br{\frac{\kappa_{\infty}}{{\epsilon_{\infty}}^{\frac{1}{\au{-}1}}}}^{\alpha{-1}}}^{\frac{1}{\alpha{+}1}}$ \\
\hline
\end{tabular}
\caption{\label{tab:LUDR} Independent characteristics obtained in the leading-order approximation for four key points on the level line of the accretion surface $\kappa{=}\kappa_{\infty}$ determined by the initial state  $\yfty$, $\bfty$ and $\tfty$ at $r{=}\rfty$ through constants $\epsfty$ and $\kapfty$ defined in \eqref{eq:epsfty} and \eqref{eq:kapfty}. The expansion parameter $\mytau$ involving these constants is  defined in the last column separately in each case. The labels in the first column refer, respectively, to   the leftmost turning point {\it L}, the rightmost turning point {\it R},  the maximum velocity point {\it U}, and the minimum velocity point {\it D}.}
\end{table*}   

\begin{table*}[]
\centering
\begin{tabular}{|c|c||c|ccccccc|c|}
$\upsilon$ & $\alpha$  & & L & U &C & D & $\text{P}'_{\infty}$ & $\text{P}_{\infty}$ & R & $\kapfty/\kc$\\
 \hline\hline
   &   & $r/\rfty$ & $3 .58{\cdot}10^{-12}$ & $9 .46{\cdot}10^{-11}$ & $0 .135$ & $0 .223$ & $1 .00$ & $1 .00$ & $1 .00$ &   \\   &   & $v_r/c$ & $0 .0363$ & $0 .0656$ & $1.00{\cdot}10^{-2}$ & $2 .79{\cdot}10^{-11}$ & $9 .67{\cdot}10^{-5}$ & $2 .00{\cdot}10^{-5}$ & $4 .99{\cdot}10^{-5}$ &   \\ $10^{-4}$ & $5/3$ & $c_s/c$ & $0 .0363$ & $1.00{\cdot}10^{-2}$ & $1.00{\cdot}10^{-2}$ & $1.00{\cdot}10^{-2}$ & $4 .01{\cdot}10^{-5}$ & $6 .78{\cdot}10^{-5}$ & $4 .99{\cdot}10^{-5}$ & $4 .60{\cdot}10^{-9}$ \\   &   & $n/\nfty$ & $1 .54{\cdot}10^{8}$ & $3 .21{\cdot}10^{6}$ & $3 .21{\cdot}10^{6}$ & $3 .21{\cdot}10^{6}$ & $0 .207$ & $1 .00$ & $0 .400$ &   \\   &   & $T_p[K]$ & $8 .62{\cdot}10^{9}$ & $6 .53{\cdot}10^{8}$ & $6 .53{\cdot}10^{8}$ & $6 .53{\cdot}10^{8}$ & $1 .05{\cdot}10^{4}$ & $3 .00{\cdot}10^{4}$ & $1 .63{\cdot}10^{4}$ &   \\ 
\hline
  &   & $r/\rfty$ & $6 .34{\cdot}10^{-10}$ & $6 .79{\cdot}10^{-8}$ & $0 .0322$ & $0 .0530$ & $1 .00$ & $1 .00$ & $1 .02$ &   \\   &   & $v_r/c$ & $0 .0776$ & $0 .164$ & $0.
0316$ & $2 .12{\cdot}10^{-7}$ & $8 .33{\cdot}10^{-3}$ & $1 .00{\cdot}10^{-3}$ & $3 .79{\cdot}10^{-3}$ &   \\ $10^{-3}$ & $4/3$ & $c_s/c$ & $0 .0776$ & $0 .0316$ & $0 .0316$ & $0 .0316$ & $3 .33{\cdot}10^{-3}$ & $4 .74{\cdot}10^{-3}$ & $3 .79{\cdot}10^{-3}$ & $1 .11{\cdot}10^{-5}$ \\   &   & $n/\nfty$ & $2 .03{\cdot}10^{7}$ & $8 .88{\cdot}10^{4}$ & $8 .88{\cdot}10^{4}$ & $8 .88{\cdot}10^{4}$ & $0 .120$ & $1 .00$ & $0 .259$ &   \\   &   & $T_e[K]$ & $2 .73{\cdot}10^{7}$ & $4 .46{\cdot}10^{6}$ & $4 .46{\cdot}10^{6}$ & $4 .46{\cdot}10^{6}$ & $4 .93{\cdot}10^{4}$ & $1 .00{\cdot}10^{5}$ & $6 .38{\cdot}10^{4}$ &   \\ 
\hline
 &   & $r/\rfty$ & $0 .0469$ & $0 .154$ & $0 .653$ & $1 .05$ & $1 .00$ & $1 .00$
 & $3 .19$ &   \\   &   & $v_r/c$ & $0 .421$ & $0 .570$ & $0 .315$ & $0 .1000$ & $0 .473$ & $0 .100$ & $0 .210$ &   \\ $10^{-1}$ & $3/2$ & $c_s/c$ & $0 .421$ & $0 .315$ & $0 .315$ & $0 .315$ & $0 .222$ & $0 .318$ & $0 .210$ & $0 .487$ \\   &   & $n/\nfty$ & $4 .74$ & $0 .955$ & $0 .955$ & $0 .955$ & $0 .187$ & $1 .00$ & $0 .145$ &   \\   &   & $T_e[K]$ & $1 .09{\cdot}10^{9}$ & $4 .89{\cdot}10^{8}$ & $4 .89{\cdot}10^{8}$ & $4 .89{\cdot}10^{8}$ & $2 .16{\cdot}10^{8}$ & $5 .00{\cdot}10^{8}$ & $1 .91{\cdot}10^{8}$ &   \\ 
\hline
\end{tabular}
\caption{\label{tab:dane}Accretion parameters for polytropic gas described by adiabatic index $\alpha$, moving with relativistic velocities in the gravitational field of line mass density $\upsilon{\cdot} \frac{c^2}{2G}$. The parameters are shown at characteristic points L U D R on three example closed contour lines of the accretion rate surface (as depicted in \figref{fig:phys2}). A given such line is passing through a pair of points $\text{P}_{\infty}$  and $\text{P}'_{\infty}$ representing initial states (defined by the indicated radial flow velocity, the indicated temperature, and referred to some arbitrary number density $\nfty$) at some arbitrary radius $\rfty$, respectively,  $\text{P}_{\infty}$ on the subsonic branch  of the contour line and $\text{P}'_{\infty}$ on the supersonic branch of the contour line. The parameters can be compared with corresponding values at the stationary point $C$ of the accretion rate surface. The last column shows the ratio of the actual accretion rate $\kapfty$ (which is a constant on a given contour line) to the critical accretion rate $\kc$ at the stationary point $C$. The subscript of symbol $T$ indicates that the corresponding temperature was calculated assuming electron mass ($T_e$) or proton mass ($T_p$).  }
\end{table*}

The coefficients in the series expansions can be calculated term by term, however cannot be given in general form. The expansion parameter is in each case some positive power of the following characteristic combination of the initial data (proportional to $(\kapfty/\kc)^{\alpha{-}1}$)
$$t\equiv\theta_{\infty}{\cdot}\sq{\frac{\kappa_{\infty}^{\au{-}1}}{\epsilon_{\infty}}}^{\frac{\alpha{-}1}{\au{-}1}}< \theta_{\infty}\beta_{\infty}^{\alpha-1}<\theta_{\infty}.$$  The magnitude of $t$ is determined mainly by the initial temperature $\theta_{\infty}$, which often will be quite a small number. Having found the series, the other derived quantities can be expanded in the same small parameter. 

The leading order expansion terms of various observables at the four characteristic points L, U, D and R on the level lines of the accretion solutions are listed in \tabref{tab:LUDR}. The higher 
expansion terms were not shown as they get complicated and are not needed for the 
presented analysis (however, they were helpful to set the starting values for the root refinement procedure used  in the preparation of the numerical results presented below). 

Accretion parameters at characteristic points of example solutions are shown in \tabref{tab:dane}. For the initial state possible for the electron gas $\bfty{=}10^{-3}$,   $T{=}10^5$K, $\alpha{=}4/3$, in moderately strong gravitational field, $\upsilon{=}10^{-3}$,
the accretion rate is considerably lower than the critical one: $\kapfty/\kc{\sim}10^{-5}$. 
The solution domain is wide -- the interior shock position of $6{\cdot}10^{-10}\rfty$ is extremely low compared with the initial radius. 
For a cooler proton gas with lower initial velocity in a weaker field -- $\bfty{=}2{\cdot}10^{-5}$, $\tfty{=}3{\cdot}10^4$K, $\alpha{=}5/3$,  $\upsilon{=}10^{-4}$ -- the ratio $\kapfty/\kc$ decreases to $5{\cdot}10^{-9}$ and the interior shock position $4{\cdot}10^{-12}\rfty$ is practically at the center.
In the ultrarelativistic regime of electron gas ($\bfty{=}10^{-1}$, $\tfty{=}5{\cdot}10^8$K, $\alpha{=}3/2$, $\upsilon{=}10^{-1}$),  the ratio $\kapfty/\kc$ increases (the accretion domain shrinks), however solutions are still possible in a wide region.  Thus, for more realistic parameters, the $\kapfty/\kc$ ratio 
decreases and the accretion region expands. This behavior is clear from the asymptotic expansions presented in \tabref{tab:LUDR} -- the ratio of bounding radii $r_L/r_R\propto t^{\sfrac{(\au{-}1)}{\upsilon(\alpha{-}\au)}}$ goes to $0$ with $t$ (which occurs when $\kapfty{\ll}\kc$) and then also $r_R{\sim}\epsfty^{1/\upsilon}\rfty{>}\rfty$.

As can be seen from the above examples, both for hotter and cooler gas, with lower or grater initial velocity, the accretion solutions are possible for a range of radii large enough for that the solutions could model a physical process. Only for solutions with high accretion rate, close to the critical one, the solution domain shrinks to physically unacceptable narrow interval of radii.  

\subsection{Comparison with the Michel spherical model}

\newcommand{\m}{\mathfrak{m}}
In Schwarzschild metric (with areal radius $x{=}r/r_{\infty}$ and mass 
$\m{=}GM/r_{\infty}c^2$), the equations  analogous to \eqref{eq:epsfty} and \eqref{eq:kapfty} read
\begin{equation*}
\!\;\!\:\br{1{+}\frac{\alpha\,\theta_{\infty}}{\alpha{-}1}{\ydens}^{\alpha{-}1}}\sqrt{\frac{1{-}\frac{2\m}{x}}{1{-}\beta^2}}
\!=\!\epsilon,\quad \\ \quad
{x^{2}{\ydens}\,\beta}\sqrt{\frac{1{-}\frac{2\m}{x}}{1{-}\beta^2}}=\mathcal{\kappa}.
\end{equation*} Here, the convention is adopted that $x{=}1$ corresponds to some arbitrary radius $r_{\infty}$ where the initial data $n_{\infty}$, $\theta_{\infty}{=}\frac{k_BT_{\infty}}{mc^2}$ and $\beta_{\infty}$ are set (${\ydens}{=}n/\nfty$ and  ${\ydens}{=}1$ at $r{=}\rfty$).\footnote{On introducing the notation $\tilde{u}{\equiv}-\beta\sqrt{\frac{1{-}\frac{2\m}{x}}{1{-}\beta^2}}$ (the radial component of the fourvelocity vector) and the definitions of $n,e,p$ as given just before  \secref{sec:chpts}, these equations reduce to a form 
$$\br{\frac{p+e}{mc^2n}}^2\br{1{-}\frac{2\m}{x}+\tilde{u}^2}=\epsilon^2,\qquad x^2 \tilde{u} {\ydens}=\kappa$$ equivalent to one given by \citet{1972Ap&SS..15..153M}.}
Given an $\epsilon$, the critical point $(\xc,\bc)$ of the accretion surface is determined by solving a cubic equation for $\bc^2$ obtained by taking the square of the second equation below in the region $0{<}\bc{<}\sqrt{\alpha{-}1}$: 
$$x_c{=}\frac{\m}{2\bc^2}\br{1{+}3\bc^2}>2\m,
\quad 
\epsilon{=}\frac{1}{
\br{\!
1{-}\frac{\bc^2}{\alpha{-}1}
\!}
\!\!\sqrt{1{+}3\bc^2}
}{\equiv} w(\bc^2).$$
A calculation similar to one performed in \secref{sec:shocks} shows that  the sign of the Hessian determinant is $\sgn(3\alpha{-}5{-}9\bc^2)$, while $\sgn({\partial^2_{\beta}\kappa}){=}{-}\sgn(1{+}\alpha{-}3\bc^2)$. The critical point could be thus elliptic for $5{-}3\alpha{+}9\bc^2{<}0$ (that is, for $\alpha$ large enough -- at least $\alpha{>}5/3$) and since then $1{+}\alpha{>}3\bc^2$ the point would be a local maximum. Below $\alpha{=}5/3$ only hyperbolic points are  possible (if $\epsilon{>}1$) which is a situation considered by \citet{1972Ap&SS..15..153M} (the spherical accretion solutions with elliptic critical point were of no interest therein). By examining  a function of $\bc$ defined on the right of the equality sign in the second equation above, one can infer that for an elliptic point to occur one needs $\frac{3(\alpha{-}1)}{2\br{\alpha{-}2/3}^{3/2}}{<}\epsilon{<}1$ and $5/3{<}\alpha{<}14/3$ (the latter condition ensures that the function has a minimum and it is located at $\beta_c{<}1$). It is possible that the presence of an elliptic point will be associated with the presence of hyperbolic point on the same phase diagram (examples of accretion phase diagrams with two critical points are known \citet{Bratek_2019}). This will not happen if additionally to previous condition also $1{>}\epsilon{>}\frac{\alpha{-}1}{2(\alpha{-}2)}$ is satisfied, which requires $\alpha{>}3$ at least, thus beyond the Taub limit. Above $\alpha{=}14/3$ there are no critical points. 

\subsection{Remarks concerning the presence of turning points}

The basic qualitative difference between spherically symmetric radial accretion on a point-like source and cylindrically symmetric radial accretion on a string-like source lies in the fact that for the latter kind of accretion the presence of turning shock points is generic. Namely, unlike for spherical model, under cylindrical symmetry the turning shock points are present for accretion rates lower than the critical value -- the critical value of accretion rate delineates the regime with no solutions from that with solutions present in between shock cylinders.

For Bondi spherical accretion the infall solutions from a large distance with accretion rates exceeding the critical one are possible and they break off at turning shock points at radii above the critical radius. This is different from the present cylindrical model in which the critical accretion rate is the maximum above which solutions are not possible - in the spherical model the critical point is hyperbolic for $\alpha{<}5/3$ and in the cylindrical model it is elliptic. The advantageous property of the cylindrical model  is that the accretion rates are substantially lower than critical values for a broad range of values characterizing the matter and the gravitational field.

The presence of turning shock points in this and other accretion models not necessarily has to be considered as nonphysical. Accretion models admitting sufficiently extended solution domains between turning points with a size corresponding to realistic astrophysical situations can be considered as physically viable. This is the case for the present model  -- for a wide range of initial data, the accretion rate characterizing the solutions is much lower than the 
critical one. This is the most important feature of the model, decisive for its practical
suitability. The model might be useful in understanding real accretion processes. Furthermore, the presence of turning shock points only signalizes breaking off a continuous branch of an idealized solution. These features of idealized models will probably be removed by taking into account dissipative processes or the interaction with magnetic fields.

\section{Summary}

Investigated in this work were  purely radial accretion solutions in the Levi-Civita space-time of a non-rotating string. Because of the form of the equations governing the accretion, one could limit the analysis to a special Wilson form of the metric (the simplification is possible or not possible depending on particular observables of interest, which was illustrated on the example of the redshift formula and on the example of the Kepler problem).  The accretion model can be regarded as a cylindrically-symmetric counterpart of the spherically-symmetric radial accretion in Schwarzschild spacetime \cite{1972Ap&SS..15..153M}. Instead of trying to find exact solutions, the solutions were studied qualitatively as isocontours of the accretion rate surface (and compared in a particular case with the exact solution). Where necessary, the asymptotics of the solutions at the symmetry axis was given. The solutions were parameterized with two dimensionless constants: $\upsilon$ (measuring the mass per length of the string in units of $\frac{c^2}{2G}\,{\sim}6.73{\times}\,10^{27}\mathrm{g}{\cdot}\mathrm{cm}^{-1}$) and $\alpha$ (the adiabatic index of the considered equation of state). There are three kinds of phase diagrams depending on whether $\alpha$ is lower, equal or grater than the critical adiabatic index $\au {\equiv} 1{+}\frac{\upsilon\left( 1{ +} \upsilon  \right)  }{1 {+} \upsilon  {+} {\upsilon }^2}{<}\,5/3$. There are no solutions extending to the spatial infinity. All solutions consist of a subsonic and a supersonic branch. The solutions are spatially bounded by two shock sonic points at which the two branches join with each other if $\alpha{>}\au$ (in which case the accretion rate is bounded from above by a critical accretion rate), or by a single shock point and the symmetry axis $x{=}0$ if $\alpha{=}\au$ (with finite velocity at $x{=}0$, respectively, $0{<}\beta{<}\sqrt{\au{-}1}$ for the subsonic or $\sqrt{\au{-}1}{<}\beta{<}1$ for the supersonic branch), or for  $\alpha{\leqslant}\au$ with the velocity attaining $0$ or $1$ at $x{=}0$ (in the latter case the accretion rate can be arbitrary high). Although the accretion solutions are not known in exact form, the physical analysis was possible approximately by finding characteristic points on the contour line of the accretion rate surface, where the physical parameters of the accreting gas could be found (in the leading order of generalized power series expansion method and then refined by numerical methods). Three examples of accretion in different physical conditions were presented. The obtained results seem physically viable -- the solution domain is large compared to the initial or maximum radius both for moderately weak and for strong fields and reasonable physical conditions of the accretion (temperatures and flow velocity), only for accretion rates close to the critical the solution domain shrinks below physically acceptable limits.  
  
\bibliographystyle{apj}
\bibliography{literatura} 
 
\end{document}